\documentclass[aps,pra,twocolumn,superscriptaddress,showpacs]{revtex4-1}

\usepackage{braket} 
\usepackage[utf8]{inputenc}
\usepackage[english]{babel}
\usepackage{graphicx}
\usepackage{amsmath}
\usepackage{lipsum} 
\usepackage[hidelinks]{hyperref}
\usepackage{mathtools}
\usepackage{amssymb}
\usepackage{lmodern} 
\usepackage[T1]{fontenc}
\def\beq{\begin{equation}}
\def\eeq{\end{equation}}
\def\ba{\begin{align}}
\def\enda{\end{align}}
\def\bi{\begin{itemize}}
\def\ei{\end{itemize}}

\def\ket#1{|#1\rangle}

\DeclarePairedDelimiter\floor{\lfloor}{\rfloor}

\usepackage{ulem}
\usepackage[usenames,dvipsnames]{color}

\usepackage{tabularx}

\begin{document}

\title[]{Superfluid phases induced by the dipolar interactions}

\author{Rebecca Kraus}
\affiliation{Theoretical Physics, Saarland University, Campus E2.6, D--66123 Saarbr\"ucken, Germany}
\author{Krzysztof Biedro\'n}
\affiliation{Institute of Theoretical Physics, Jagiellonian University in Krakow, ul. Lojasiewicza 11,
30-348 Krak\'ow, Poland}
\author{Jakub Zakrzewski}
\affiliation{Institute of Theoretical Physics, Jagiellonian University in Krakow, ul. Lojasiewicza 11,
30-348 Krak\'ow, Poland}
\affiliation{Mark Kac Complex Systems Research Center,  Jagiellonian University in Krakow, \L{}ojasiewicza 11, 30-348 Krak\'ow, Poland}
\author{Giovanna Morigi}
\affiliation{Theoretical Physics, Saarland University, Campus E2.6, D--66123 Saarbr\"ucken, Germany}
\date{\today} 

\begin{abstract}
We determine the quantum ground state of dipolar bosons in a quasi one-dimensional optical lattice and interacting via $s$-wave scattering. The Hamiltonian is an extended Bose-Hubbard model which includes hopping terms due to the interactions. We identify the parameter regime for which the coefficients of the interaction-induced hopping terms become negative. For these parameters we numerically determine the phase diagram for a canonical ensemble and by means of density matrix renormalization group. We show that at sufficiently large values of the dipolar strength there is a quantum interference between the tunneling due to single-particle effects and the one due to the interactions. Because of this phenomenon, incompressible phases appear at relatively large values of the single-particle tunneling rates. This quantum interference cuts the phase diagram into two different, disconnected superfluid phases. In particular, at vanishing kinetic energy the phase is always superfluid with a staggered superfluid order parameter. These dynamics emerge from quantum interference phenomena between quantum fluctuations and interactions and shed light into their role in determining the thermodynamic properties of quantum matter.\end{abstract}

\maketitle

\section{Introduction}
Ultracold atoms in optical lattices are prominent platforms that shed light on the interplay between interactions and quantum fluctuations. That interplay determines key properties of 
quantum matter \cite{RMP:Bloch-Dalibard-Zwerger}. In fact, in these systems it is possible to experimentally tune the relative strength of quantum fluctuations and of interactions. This capability enables one, for instance, to sweep across the superfluid-Mott insulator quantum phase transition in a gas of bosons \cite{Fisher1989,Greiner2002}, just to mention a remarkable example. In this scenario, the experimental observation of quantum phases of ultracold dipolar gases \cite{Tanzi:2019,Chomaz2019,Boettcher2019} and their confinement in optical lattices  \cite{Je,Baier2016,dePaz:2013,Moses:2015,Covey:2016,Reichsoellner:2017, Lahaye} paves the way towards the characterization of strongly correlated quantum matter, which is typically theoretically described by the so-called extended Hubbard model \cite{Pollet:2010,Lahaye,Pupillo,Dutta2015,Menotti2007,Goral2002,Yi:2007,Pupillo:2010}.  


In the extended Bose-Hubbard model the effect of power-law interactions is usually represented by density-density interaction terms \cite{Pupillo,Lahaye,Dutta2015, Menotti2007,Goral2002,Yi:2007,Pupillo:2010,Pollet:2010,Sinha:2005}. These terms are responsible for density modulations within the lattice \cite{Dutta2015,Menotti2007,Goral2002,Pupillo:2010,Yi:2007, Sinha:2005,Otterlo:1994,Mishra:2009,Sengupta2005,Batrouni2006,Kuehner:1999,Batrouni1995} and for topological incompressible phases in one dimension \cite{DallaTorre2006,Deng2011,Rossini2012,Batrouni2013,Kawaki:2017}. Moreover, the onsite contribution of the dipolar potential typically renormalizes the contact interactions and can make the gas unstable \cite{Santos:2000,Goral2002,cartarius:2017,Goral:2002}.  {\it Ab initio} derivations of the Bose-Hubbard model show that interactions are also responsible for the appearance of correlated hopping terms \cite{Schmidt}, which can be of the same order as the density-density interactions terms \cite{sowinski2012dipol,Baier2016,cartarius:2017,biedron2018extended,Amico:2010}. 

There is a limited knowledge on the effect of the interaction-induced hopping on the ground state properties of a dipolar gas in an optical lattice. Exact diagonalization for a chain of few bosons showed the appearance of exotic superfluid and charge-density wave phases \cite{sowinski2012dipol}. Density matrix renormalization group (DMRG) studies of the extended grandcanonical Bose-Hubbard model found superfluid phases with non-vanishing Fourier components which can be either commensurate or incommensurate with the lattice periodicity, depending on the lattice depth \cite{biedron2018extended}. These studies point out that the interplay between the various quantities is by no means trivial and calls for a systematic study of the phases as a function of the power-law interaction strength. 

In this paper we study the extended Bose-Hubbard model of a quasi-one dimensional gas of ultracold dipolar bosons. We identify the parameter regime where the interaction-induced hopping terms become of the same order as the kinetic energy and determine the resulting ground state phase diagram. The phase diagram is numerically determined for a canonical ensemble at zero temperature and at density $\rho=2$ per lattice site by means of a DMRG program  \cite{Schollwoeck2011,itensor}, the phases are characterized using the classification discussed in \cite{Johnstone19}. We find several remarkable properties. Among them, the most striking is that correlated hopping can destructively interfere with the hopping due to the kinetic energy. This quantum interference is responsible for the appearance of incompressible phases in relatively shallow lattices. The interaction induced hopping, moreover, gives rise to superfluidity in deep lattices, where otherwise one would expect incompressible phases. These superfluid phases are characterized by a staggered superfluid order parameter.

The content of this paper is here summarized. In Sec. \ref{Sec:2} we introduce the Bose-Hubbard model for power-law interacting atoms and discuss the behavior of the coefficients for dipolar interactions. In Sec. \ref{Sec:3} we analyze the phases at zero temperature by means of a mean-field ansatz. The results of the numerical simulations are reported in Sec. \ref{Sec:4} and the conclusions are drawn in Sec. \ref{Sec:5}. The appendices provide details of the calculations of the coefficients, of the numerical implementations, as well as supplementary results.


\section{Extended Bose-Hubbard model}
\label{Sec:2}

We consider ultracold bosons of mass $m$ in an anisotropic trap which is elongated along the $x$ axis. The dipolar bosons are polarized by an external field perpendicular to the trap axis and interact via the dipolar and the van-der Waals ($s$-wave) interactions. 
The dynamics is governed by the second-quantized Hamiltonian for the bosonic field $\hat{\Psi}(\mathbf{r})$ \cite{cartarius:2017}:
\begin{align}
\hat{H}=&\int d^3\mathbf{r} \hat{\Psi}^\dagger (\mathbf{r})\left[-\frac{\hbar^2}{2m}\nabla^2+V_{\text{trap}}(\mathbf{r})\right]\hat{\Psi}(\mathbf{r})\nonumber\\
&+\frac{1}{2}\int d^3\mathbf{r} \int d^3\mathbf{r}'\hat{\Psi}^\dagger(\mathbf{r})\hat{\Psi}^\dagger (\mathbf{r}')U_{\rm int}(\mathbf{r}-\mathbf{r}')\hat{\Psi}(\mathbf{r}')\hat{\Psi}(\mathbf{r}) \ ,  \label{secquant}
\end{align}
where the field operators $\hat{\Psi}(\mathbf{r})$ and $\hat{\Psi}^\dagger(\mathbf{r})$ obey the commutation relation $\left[\hat{\Psi}(\mathbf{r}), \hat{\Psi}(\mathbf{r}')^\dagger \right]=\delta ^3\left(\mathbf{r}-\mathbf{r}' \right)$. The function $V_{\text{trap}}$ denotes the trap potential, which is the sum of a tight harmonic potential in the $y$-$z$ plane, and an optical lattice of periodicity $a$ along the $x$ axis, 
\begin{equation}
 V_{\text{trap}}=\frac{m\omega^2}{2}\left(y^2+ z^2\right) +V_0\sin^2(\pi x /a)\,.
 \label{V:trap}
 \end{equation}
Here, $\omega$ is a harmonic trap frequency and $V_0$ denotes the amplitude of the optical lattice. 
The interaction potential is the sum of the contact and of the power-law interactions,
\begin{equation}U_{\rm int}(\mathbf{r})=U_g(\mathbf{r})+U_\alpha(\mathbf{r})\,. 
\label{eq:potential}
\end{equation}
Specifically, $U_g(\mathbf{r})=g\delta^{(3)}(\mathbf{r})$ is the contact potential  with $g=4\pi \hbar ^2 a_s/m$ and $a_s$ the $s$-wave scattering length. The power-law interactions, $U_\alpha(\mathbf{r})$,  scale with the interparticle distance $r$ as $U_\alpha(\mathbf{r})\propto 1/r^\alpha$. In this work we consider dipoles polarized by an external field along the $z$ axis. In this case $U_\alpha(\mathbf{r})\equiv U_d(\mathbf{r})$, where
\begin{align}
U_d(\mathbf{r})=\frac{C_{dd}}{4\pi} \frac{1-3\cos^2(\theta)}{r^3} \,,
\end{align}
where $\theta$  is the angle between the dipole and $\mathbf{r}$.
The dipole-dipole interaction is anisotropic in space since the force depends on the dipoles orientation.
The coefficient $C_{dd}$ scales the strength of the dipole-dipole interactions: for magnetic dipoles with moment $\mu_m$, the coefficient is $C_{dd}=\mu_0\mu_m^2$; for electric dipoles with moment $\mu_e$ it reads as $C_{dd}=\mu_e^2\epsilon_0$, where $\mu_0$ and $\epsilon_0$ are the magnetic and the electric permeability, respectively. In the rest of this paper, in place of $C_{dd}$ we will use the rescaled, dimensionless quantity $d$, that is defined as \cite{Astrakharchik2007,biedron2018extended}
\begin{eqnarray}
\label{Eq:d}
d=\frac{mC_{dd}}{2\pi^3\hbar^2a}\,.
\end{eqnarray}

\subsection{Extended Bose-Hubbard Hamiltonian}

We assume that the bosons are in the ground state of the harmonic trap and tightly bound at the minima of the optical lattice. In this low energy limit the field operator $\hat \Psi(\mathbf{r})$ is decomposed into the sum of bosonic operators $\hat{a}_{j}$, which annihilate a particle at the $j$-th lattice site \cite{Jaksch:1999}: 
\begin{align}\label{exp}
\hat \Psi(\mathbf{r})=\sum_{j=1}^L \phi_0(y,z)\omega_j(x)\, \hat{a}_{j}\,,
\end{align} 
where $j=1,\ldots,L$ and $L$ is the number of lattice sites. The scalar function $\phi_0(y,z)$ denotes the ground state of the transverse harmonic trap, $w_j(x)$ is the real-valued Wannier function for the lowest lattice band, and operators $\hat{a}_{j}$, $\hat{a}_{j}^\dagger$ fulfill the commutation relation $\left[\hat{a}_{j}, \hat{a}_{l}^\dagger\right]=\delta_{j,l}$. After using Eq. \eqref{exp} in Hamiltonian \eqref{secquant} and integrating out the spatial degrees of freedom one obtains the extended Bose-Hubbard model of Refs. \cite{sowinski2012dipol,cartarius:2017}. A summary of the derivation is reported in Appendix \ref{appendix:1}.

We denote by $\hat{H}_{\text{BH}}$ the corresponding Bose-Hubbard model, and decompose it into the sum of the Hamiltonian $\hat H_0$, describing the onsite interactions and tunneling term, and of the Hamiltonian $\hat H_\alpha$, which includes the other terms. Hamiltonian $\hat H_{\rm BH}^{(0)}$ takes the form of the "traditional" Bose-Hubbard model \cite{Fisher1989}:
\begin{eqnarray}
\label{H:BH}
\hat H_{\rm BH}^{(0)}=-t\sum_{j=1}^{L-1}\left(\hat{a}^\dagger_j\hat{a}_{j+1}+{\rm H.c.}\right) +\frac{U}{2}\sum_{j=1}^L \hat{n}_j\left(\hat{n}_j-1\right)\,
\end{eqnarray}
where $\hat n_j=\hat a_j^\dagger \hat a_j$ counts the number of particles at site $j$. The coefficients $t$ and $U$ denote the tunneling amplitude and the onsite interaction, respectively. Coefficient $t$ solely depends on expectation values of the single-particle Hamiltonian, and in particular on the kinetic energy. The onsite interaction coefficient $U$, instead, is determined by the corresponding integral of the potential of Eq. \eqref{eq:potential} (see Appendix \ref{appendix:1}). Thus, in general it also includes the contribution of the power-law interactions. {Depending on the trap geometry, this contribution can lead to vanishing or negative values of the onsite interactions, which may cause a collapse of the  system \cite{Santos:2000,Goral:2002,Goral2002,sowinski2012dipol,cartarius:2017}. In this work we will restrict ourselves to geometries for which the coefficient $U$ is positive, thus the onsite interactions are repulsive and the gas is stable.

The Bose-Hubbard Hamiltonian $\hat H_{\rm BH}^{(0)}$ is obtained (i) by truncating the hopping processes to the nearest neighbors and (ii) by solely taking the local contribution of the interactions. Deep in the tight-binding regime the first approximation is justified. On the contrary, for large values of the onsite potentials and/or in the presence of power-law interactions one shall consistently include the coupling between $\ell$-th nearest neighbor. These terms are contained in the Hamiltonian $\hat H_\alpha$, which we write as the sum of the terms coupling  $\ell$-th nearest neighbor:
$$\hat H_\alpha=\sum_{\ell=1}^L\hat{H}_{\alpha}^{(\ell)}\,,$$
and whose detailed form is given below for $\ell=1,2$.

In this work we truncate the sum over $\ell$ and analyze the phase diagrams of the Hamiltonian for two cases. First we consider the ground state of the Bose-Hubbard Hamiltonian, where we truncate the power-law interactions to the nearest-neighbors:
\begin{equation}
\label{HBH:nn}
\hat{H}_{\text{BH}}^{(1)}=\hat{H}_{\text{BH}}^{(0)}+\hat H_{\alpha}^{(1)}\,.
\end{equation}
We then compare the corresponding phase diagrams with the ones obtained by keeping also the coupling to the next-nearest neighbors:
\begin{equation}
\label{HBH:nnn}
\hat{H}_{\text{BH}}^{(2)}=\hat{H}_{\text{BH}}^{(1)}+\hat H_{\text{NNN}}+\hat H_{\alpha}^{(2)}\,.
\end{equation}
Here, $\hat H_{\text{NNN}}$ describes the next-nearest neighbor hopping terms due to the kinetic energy and to the trapping potential, which we include for consistency(see Appendix \ref{appendix:1}).  

In the rest of this section we discuss the detailed form of $\hat{H}_{\alpha}^{(1)}$ and of $\hat{H}_{\alpha}^{(2)}$.
The Hamiltonian $\hat{H}_{\alpha}^{(1)}$ reads as \cite{sowinski2012dipol,jakub2013,Johnstone19}
\begin{align}
\hat{H}_{\alpha}^{(1)}=&V\sum_{j=1}^{L-1}\hat{n}_{j}\hat{n}_{j+1}-T\sum_{j=1}^{L-1}\left[\hat{a}^\dagger_j\left( \hat{n}_j+\hat{n}_{j+1}\right)\hat{a}_{j+1}+{\rm H.c.}\right]\nonumber \\
&+\frac{P}{2} \sum_{j=1}^{L-1}\left(\hat{a}^\dagger_{j+1}\hat{a}^\dagger_{j+1} \hat{a}_j\hat{a}_j +{\rm H.c.}\right)\,. \label{BHHnn}
\end{align}
The term scaled by the positive amplitude $V$ describes a repulsive density-density interaction. This term tends to inhibit the occupation of neighboring sites and promotes density modulations. In the following we denote it by blockade coefficient. The other two terms describe tunneling effects induced by the interactions. In detail, coefficient $T$ scales a hopping term which nonlinearly depends on the occupation number of neighboring sites. We will denote this term by "density-assisted tunneling". The term scaled by $P$ describes pair hopping between nearest-neighbors and we will refer to it as "pair-hopping term". 

The form of the higher-order coupling terms is similar to the one of $\hat{H}_{\alpha}^{(1)}$. We report here the coupling to next-nearest neighbors:
\begin{align}
\hat{H}_{\alpha}^{(2)}=&V_{\text{NNN}}\sum_{j=1}^{L-2}\hat{n}_{j}\hat{n}_{j+2} +\sum_{j=1}^{L-2}\left(\hat{T}_j^{(2)}+\hat{P}_j^{(2)}\right)\,. \label{BHHnnn}
\end{align}
Here $V_{\text{NNN}}$ scales term describing the repulsive next-nearest neighbor density-density interaction and is positive. The corresponding interaction-induced tunneling and pair hopping terms are now collected in operators $\hat T^{(2)}_j$ and $\hat{P}^{(2)}_j$, respectively, and take the form:
\begin{align}
\hat{T}_j^{(2)}&=-T_{\text{NNN}}\hat{a}^\dagger_{j}\left(\hat{n}_j+\hat{n}_{j+2}\right) \hat{a}_{j+2}
-T_{\text{NNN}}^1\hat{a}^\dagger_{j+1}\hat{n}_j\hat{a}_{j+2} \nonumber\\
&-T_{\text{NNN}}^2\hat{a}^\dagger_{j}\hat{n}_{j+2}\hat{a}_{j+1}-T_{\text{NNN}}^3\hat{a}^\dagger_{j}\hat{n}_{j+1}\hat{a}_{j+2}+{\rm H.c.}\,,\label{densitydependent}\\
\hat{P}_j^{(2)} =&\frac{P_{\text{NNN}}^1}{2}\hat{a}^\dagger_{j+1}\hat{a}^\dagger_{j+2}\hat{a}_{j}\hat{a}_{j} + \frac{P_{\text{NNN}}^2}{2}\hat{a}^\dagger_{j+2}\hat{a}^\dagger_j\hat{a}_{j+1}\hat{a}_{j+1} \nonumber \\
&+ \frac{P_{\text{NNN}}^3}{2}\hat{a}^\dagger_{j+2}\hat{a}^\dagger_{j+2}\hat{a}_{j+1}\hat{a}_{j} + {\rm H.c.}  \, .\label{pairtunnelling}
\end{align}
The specific form of the coefficients $V_{\text{NNN}}$, $T_{\text{NNN}}^\ell$, $P_{\text{NNN}}^\ell$ is given in the Appendix A.

We remark that correlated hopping terms have been also discussed for atoms solely interacting via $s$-wave scattering but in the limit of large ratios $U/t$ \cite{Luehmann2012,Juergensen2014}. In this case, for repulsive interactions the coefficients are all positive and the density-dependent tunneling leads to a reduction of the incompressible region \cite{Luehmann2012}. In the next section we identify a parameter regime where there is a sign change of the interaction-induced hopping coefficients as a function of $d$.

\subsection{Interaction-induced tunneling}
\label{Sec:2.0}
 
By changing the quantum species, and thus changing $d$, one modifies the relative weight between $s$-wave and dipolar interactions. The first one typically dominates at short-range distances, while the dipolar interactions are expected to determine the non-local terms. We now consider a trap geometry, where $s$-wave scattering is negative and the onsite contribution of the dipolar interactions stabilizes the gas, making the onsite interactions repulsive, $U>0$. In this regime, we identify the parameter regime where the correlated tunneling coefficients become negative.

Figure \ref{appfig:1}(a) and (b) display the contour plots of the density-assisted tunneling coefficient $T$ and of the pair tunneling coefficient $P$ as a function of $U$ and $V$, keeping $t$ fixed. For later convenience we label the axis by $V/U$ and $t/U$. In subplot (a), moreover, we explicitly show lines at constant dipolar interaction $d$. We observe that the value of $T$ becomes comparable with $V$ at large dipolar interaction strength and for small ratios $t/U$. Therefore, when $t/U\to 0$, one still has significant hopping due to the interactions. We note, moreover, that the pair tunneling coefficients remain very small across the phase diagram. We will keep these terms in our simulations, and anticipate that they play a negligible role in determining the phases of the ground state for deep lattices. 
\begin{figure}[h!]
	\centering	
	(a)
	\begin{minipage}[t]{0.37\textwidth}\vspace{0pt}
		\includegraphics[width=\textwidth]{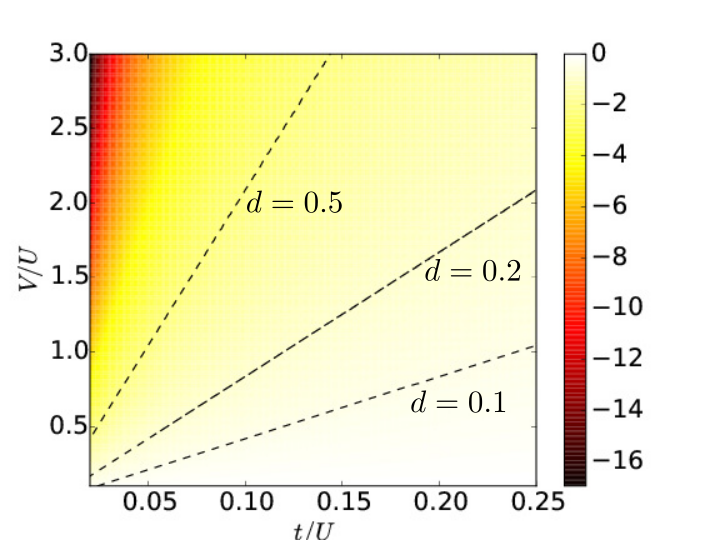}
	\end{minipage}\hfill%
	
	(b)
	\begin{minipage}[t]{0.37\textwidth}\vspace{0pt}
		\includegraphics[width=\textwidth]{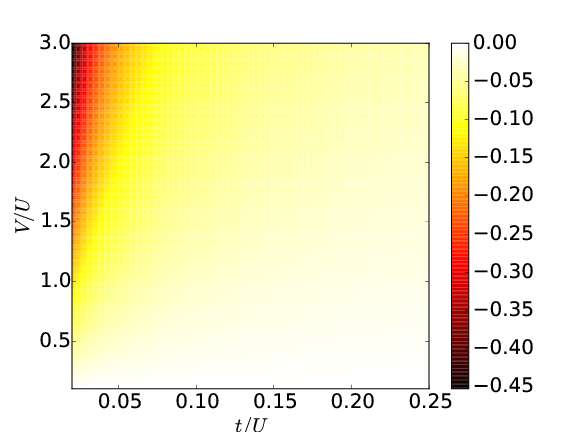}
	\end{minipage}\hfill
	\caption{(color online) Contour plot in the $V/U-t/U$-plane of the (a) density-assisted tunneling coefficient $T$ and (b) the pair tunneling coefficient $P$ for the nearest-neighbor coupling and in units of the tunneling rate $t$. The black dashed lines show the values of $V$ and $U$ at some constant dipolar interaction strengths $d$. The other parameters are discussed in the text.} 
	\label{appfig:1}
\end{figure}

Let us here specify the parameters we used for evaluating these coefficients and which we will use in the rest of this paper, unless otherwise stated. The depth of the optical lattice in the axial direction is kept fixed to the value $V_0=8E_R$, where $E_R$ is the recoil energy. The transverse trap frequency is $\omega=\sqrt{2V_{\rm har}\pi^2/a^2m}$, where we choose $V_{\rm har}=50E_R$. The tunneling rate $t$ between nearest-neighbor and the tunneling rate between next-nearest neighbor $t_{\text{NNN}}$ are constant, and for the given lattice depth $t_{\text{NNN}}=0.0123 \,t$.


\subsection{Interaction-induced atomic limit}

We now analyze the behavior of the interaction-induced tunneling coefficients as a function of $t/U$ and at a given ratio $V/U$. Figure \ref{fig:d}(a) displays the density-assisted tunneling coefficient $T$ and the pair hopping coefficient $P$ in units of $t$. The coefficients $T$ and $P$ are negative over the considered parameter range. In particular, $P$ is one order of magnitude smaller than the density dependent tunneling coefficient $T$, while $T$  is of the same order of magnitude as the tunneling rate. Therefore, single-particle tunneling and correlated tunneling have opposite sign and can mutually cancel. Correlated (density-assisted) tunneling, in particular, is dominant for $t/U \to 0$, while single-particle hopping is dominant at large ratios $t/U$. There is a parameter range at finite ratios $t/U$, thus, where this destructive interference leads to an effective atomic limit. Figure \ref{fig:d}(b) displays the values of the scattering length and of the dipolar interaction strength corresponding to the curves in subplot (a). The ratio $t/U$ for which one finds the interaction-induced atomic limit is indicated by the vertical black line in subplot (b) for density $\rho=2$ per lattice site. Here, $T=t/3$. 
\begin{figure}[h!]
	\centering
	(a)
	\begin{minipage}[t]{0.4\textwidth}\vspace{0pt}
		\includegraphics[width=\textwidth]{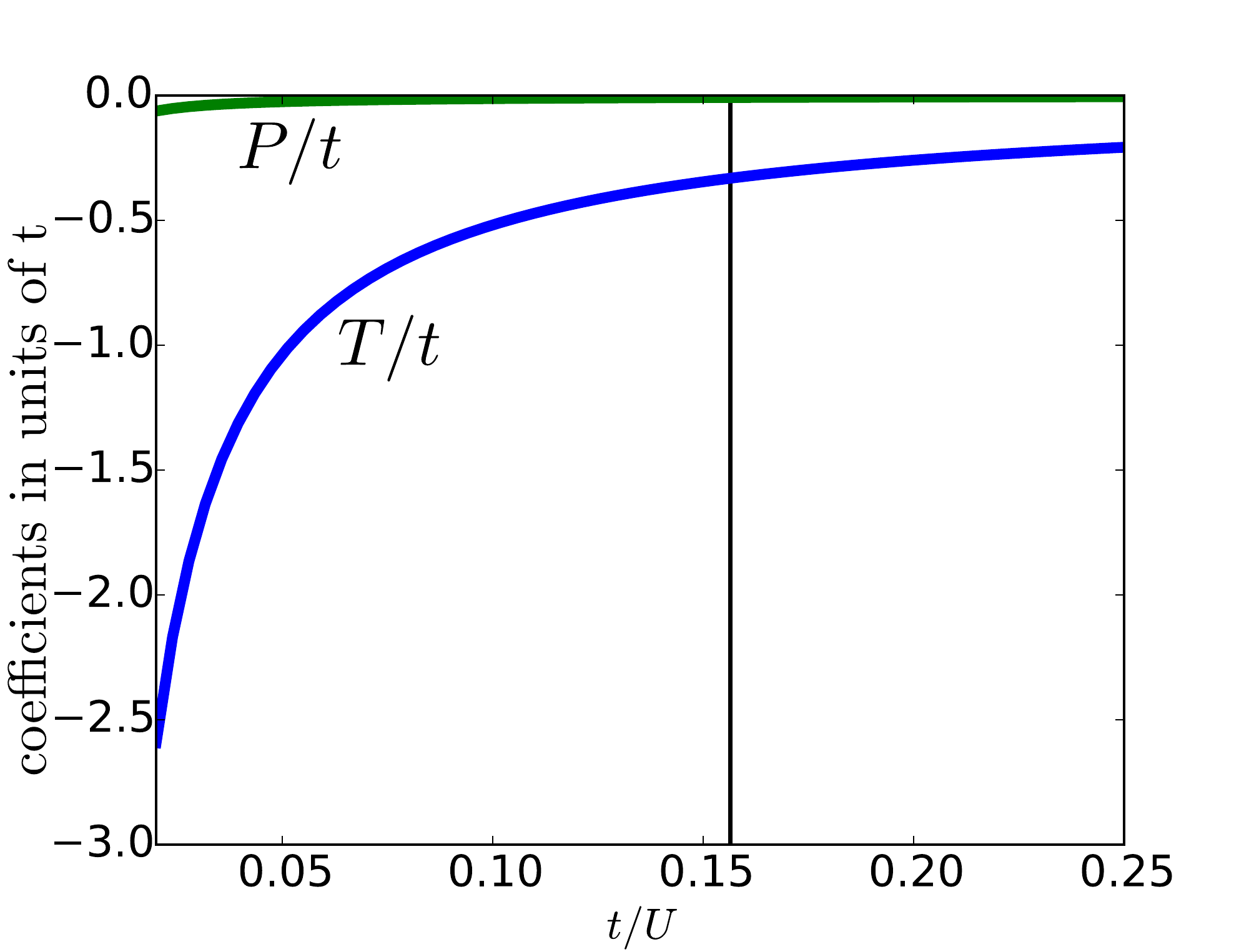}
	\end{minipage}\hfill\\
	(b)
		\begin{minipage}[t]{0.4\textwidth}\vspace{0pt}
			\includegraphics[width=\textwidth]{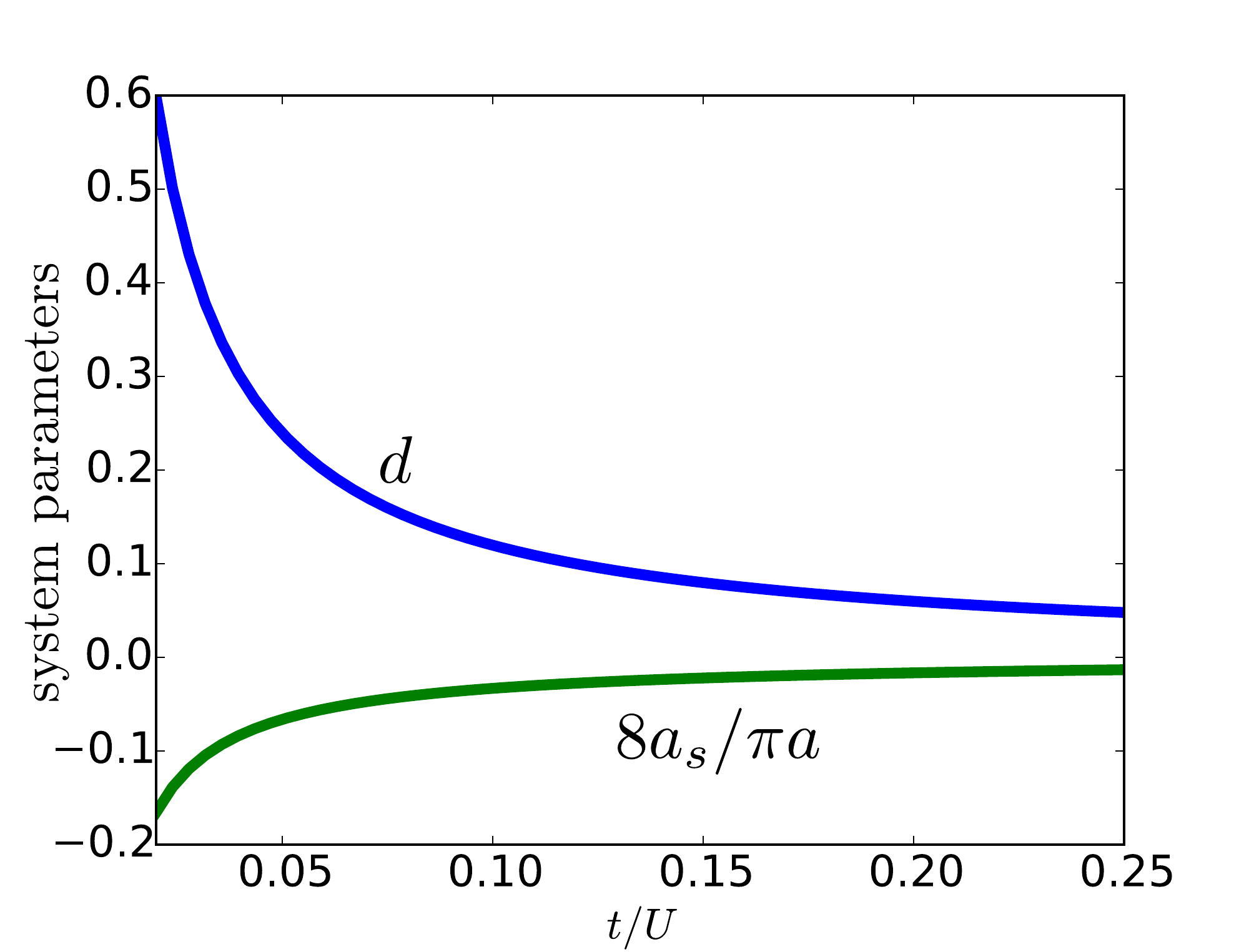}
		\end{minipage}\hfill\\
	\caption{(color online) (a) Density-assisted tunneling coefficient $T$ (blue) and pair-hopping coefficient $P$ (green) as a function of $t/U$ for $V/U=0.5$.  Subplot (b) displays the corresponding values of the dimensionless dipolar interaction strength $d$ (blue) and of the $s$-wave scattering length $a_s$ in units of $a$ (green). The ratio $V/U$ is increased by changing both $d$ and the scattering length $a_s$. Note that here we choose negative scattering lengths, so that the $s$-wave potential partially cancels out with the local repulsive component of the dipolar interactions. The other parameters are given in the text. The vertical black line in subplot (a) indicates the value of $t/U$ for which linear tunneling and density-assisted tunneling between nearest-neighbor sites mutually cancel. } 
	\label{fig:d}
\end{figure}

In order to find a systematic way to identify the regime of the interaction-induced atomic limit, we consider the one-particle hopping terms of Hamiltonian $H_{\rm BH}^{(1)}$ in Eq. \eqref{BHHnn} and collect them in the operator  $\hat {\mathcal T}_{\rm eff}$, which reads as \cite{Luehmann2012}:
\begin{equation}
 \hat {\mathcal T}_{\rm eff}=\sum_{j=1}^{L-1}\hat{a}^\dagger_j\left[-t-T(\hat n_j+\hat n_{j+1})\right]\hat{a}_{j+1} +{\rm H.c.}\,.
\end{equation}
The expectation value of this term on a homogeneous distribution at average density $\langle \hat n_i\rangle=\rho$ scales as
\begin{equation}
\langle \hat {\mathcal T}_{\rm eff}\rangle\simeq L{\rm Re}\{\langle\hat{a}^\dagger_j\hat{a}_{j+1}\rangle\} [-t-T(2\rho-1)]\,.
\end{equation}
This expression shows that, for non-vanishing off-diagonal correlations, then $\langle \hat {\mathcal T}_{\rm eff}\rangle$ can vanish when $$-t-T(2\rho-1)=0\,.$$  Solutions with $t\neq 0$ exist for densities $\rho\neq 1/2$. In particular, for $0<\rho<1/2$, destructive interference occurs for $T>t>0$. In this case, $\langle \hat {\mathcal T}_{\rm eff}\rangle=0$ for $T/t=1/(1-2\rho)$. In the other regime, for $\rho>1/2$,  the interaction-induced atomic limit requires $T<0$ and is found for 
\begin{equation}
\label{eq:interference}
\frac{|T|}t=\frac{1}{2\rho -1}\,.
\end{equation}
Relation \eqref{eq:interference} can be fulfilled at relatively small dipolar strength when the density is sufficiently high. In order to explore the effects of this interference on the ground state properties, in our numerical studies we focus on the phases of a dipolar gas with commensurate density $\rho=2$.

\section{Mean-field analysis}
\label{Sec:3}

In this Section we use a mean-field ansatz and an approximated model in order to infer some features of the phase diagrams of Sec. \ref{Sec:4}.

\subsection{Atomic limit}

We analyze the ground state in the limit in which all hopping terms are set to zero. For convenience we consider the simplified Hamiltonian
\begin{equation}
H_{\rm at}=\frac{U}{2}\sum_j\hat n_j(\hat n_j-1)+V\sum_j\sum_{r>0}\frac{1}{r^\alpha}\hat n_j\hat n_{j+r}\,,
\end{equation}
where $\alpha$ is the power law exponent, $\alpha>1$, and $V$ scales the interaction in the limit in which the Wannier functions are approximated by Dirac-delta function. 

For $\alpha\to\infty$ the interaction reduces to nearest-neighbors and the ground state results from the interplay between the onsite interaction, which tends to minimize the onsite occupation, and the repulsive interaction between neighoring sites, which favours the onset of density waves with double lattice periodicity. We denote by MI[2] the Mott-insulator state with two particle per site and by CDW$[n_1,n_2]$ the charge density wave state where neighboring sites are occupied by a repeating sequence of $n_1$ and $n_2$ particles per site. 
For $\rho=2$ these are CDW$[3,1]$ and CDW$[4,0]$. The MI[2] is stable when $V\le V_c^{(1)}$, where:
\begin{equation}
\label{MI:CDW}
V_c^{(1)}=\frac{U}{2}\frac{1}{\zeta(\alpha)\left(1-\frac{1}{2^{\alpha-1}}\right)}\,,
\end{equation}
and $\zeta(\alpha)$ is the Riemann's zeta function. For $\alpha\to\infty$ we recover the value $V_c^{(1)}=U/2$. For $\alpha$ finite the boundary is shifted to larger values of $V$: the power-law tails tend to stabilize the MI[2] state.  

For $\alpha>1$ the three phases, MI[2], CDW$[3,1]$ and CDW$[4,0]$, are degenerate at $V=V_c^{(1)}$. In the interval $V_c^{(1)}<V<V_c^{(2)}$ the ground state is the CDW[4,0]. The upper bound $V_c^{(2)}$ is given by the expression
\begin{equation}
\label{CDW:CDW}
V_c^{(2)}=\frac{2U}{\zeta(\alpha)}\frac{2^\alpha/8}{\left(1-\left(2/3\right)^\alpha/6\right)}\,,
\end{equation}
which separates the phase CDW[4,0] from the CDW$[6,0,0]$ with triatomic Wigner-Seitz cell. For repulsive dipolar interactions, when $\alpha=3$, then this bound takes the value $V_c^{(2)}\simeq 2U$. In general, $V_c^{(2)}$ monotonously increases with $\alpha$ and reaches $V_c^{(2)}\to \infty$ for $\alpha\to\infty$. For finite $\alpha$ one finds for $V>V_c^{(2)}$ further transition points to structures with increasing Wigner-Seitz cells, until all particles are localized at one lattice site in the limit $V/U\to\infty$.

\subsection{Staggered superfluidity}

For nearest-neighour interactions and $\rho>1/2$, correlated tunneling dominates the hopping events for $T<0$ and $t<|T|(2\rho-1)$. In this regime, in the absence of interactions the state with minimum energy has momentum $q=\pi$ (here in units of $1/a$). Following these considerations, we now assume a site-dependent superfluid order parameter, which we define as \cite{Juergensen2015}
\begin{equation}
\label{Ansatz}\left\langle\hat{a}_{j}\right\rangle = \phi e^{i\theta_j}\,,
\end{equation}
and use it to calculate the mean-field energy of the nearest-neighbor hopping term:
\begin{align}
H_{\rm hop}=& 2[-t+|T|(2\rho-1)] \sum_{j=1}^{L-1} \phi^2\cos(\theta_j-\theta_{j+1}) \nonumber \\ 
&+\frac{P}{2} \sum_{j=1}^{L-1}\phi^4\cos(2(\theta_j-\theta_{j+1})) \ ,
\end{align} 
where $P<0$ for the parameters of this paper. Discrete translational invariance gives $\theta_j=-j\theta$, such that $\theta_j-\theta_{j+1}=\theta$ is a constant phase increment from site to site \cite{Juergensen2015}. This ansatz shows that the energy is minimal for $\theta=\pi$ in the regime where density-assisted tunneling dominates. The SF order parameter has thus a Fourier component at $q=\pi$. The alternating sign of the local superfluid parameter leads to the denomination "staggered superfluidity" (SSF) \cite{Johnstone19}. 

We now consider the power-law behavior of the interactions, and thus the coupling to the other neighbors. For this purpose we write the single-particle hopping terms due to the single-particle tunneling and to the density-assisted tunneling in the compact form 
\begin{equation}
\hat H_{\rm hop}'=\sum_j\sum_{r>0}\left(-t_{r}-T'_r[\hat n_1,\ldots,\hat n_L]\right)\hat{a}_j^\dagger \hat{a}_{j+r}+{\rm H.c.}\,,
\end{equation}
where $T'_r[n_1,\ldots,n_L]$ is a generic function of the density distribution and scales with $1/r^\alpha$, and $t_r$ is the hopping term due to the single-particle energy, such that $t_1=t$ and $t_2=t_{\text{NNN}}$. 
For the uniform density distribution $\rho$ we make the simplifying assumptions $\langle T'_r\rangle \sim -T'[\rho]/r^\alpha$ . Using Eq. \eqref{Ansatz} we obtain the expression: 
 \begin{align}
H_{\rm hop}'\sim& 2 \phi^2\sum_{j=1}^{L-1} \sum_{r>0}\left(-t_r+\frac{T'[\rho]}{r^\alpha}\right)\cos(\theta_j-\theta_{j+r}) \, .
\end{align} 
This is the relation that the site-dependent phase $\theta_j$ shall fulfil in order to achieve the interaction-induced atomic limit}. For a shallow lattice and in the regime where interactions are dominant superfluid phases can have Fourier components that are incommensurate with the lattice periodicity \cite{biedron2018extended}. For sufficiently deep lattices, which is the case we consider in this paper, the sum can be truncated at the next-nearest neighbors. Then an approximated root of the equation $H'_{\rm hop}=0$ is found by imposing $-t+T'[\rho]=0$, where now
\begin{equation}
T'[\rho]=|T|(2\rho-1)-|T^2_{\text{NNN}}|\rho\,.
\label{Tshiftnnn}
\end{equation}
One consequence is that the coupling beyond nearest-neighbors shifts the interaction-induced atomic limit to smaller values of the ratio $t/U$. The interval where $-t+T'[\rho]>0$ is now expected to be smaller than for nearest-neighbor coupling. Here, the superfluid phase is to good approximation the staggered superfluid with $\theta_j=-j\pi$. 

\section{Ground state phase diagrams}
\label{Sec:4}

In this section we numerically determine the properties of the ground state of the extended Bose-Hubbard Hamiltonian as a function of the strength of the dipolar interactions. We choose the commensurate density $\rho=2$ and calculate the ground state when the coupling is first truncated to the nearest-neighbor and then when also the next-nearest neighbors are included. 

Our results are obtained by means of a DMRG numerical program \cite{Schollwoeck2011}, which is based on the ITensor C++ library for implementing tensor network calculations \cite{itensor}. The simulations are run for $N$ particles in a lattice with $L=N/2$ sites with open boundary conditions, for different lattice sites and for different initial states.  The interested readers are referred to Appendix \ref{App:DMRG} for details on the implementation. 

Unless mentioned otherwise, the system parameters are given in Sec. \ref{Sec:2.0}. In this parameter regime mean-field estimates predict that the atomic limit is shifted to finite values of $t/U$ due to destructive interference between single-particle hopping and correlated tunneling. 

We first review the observables, by means of which we characterize the phases. In Sec. \ref{Sec:nn} we report the phase diagram of the extended Bose-Hubbard Hamiltonian with nearest-neighbor coupling, Eq. \eqref{HBH:nn}. In Sec. \ref{Sec:nnn} we then discuss the ground state phase diagram of the Hamiltonian with next-nearest-neighbor couplings, Eq. \eqref{HBH:nnn}.

\subsection{Observables}
\label{Sec:2.1}

In this work we consider a system of atoms with a finite particle numbers at vanishing temperature. We determine the phases by means of the following observables, whose expectation values are taken over the ground state.

We identify whether a phase is compressible by means of the local variance $\Delta n_j$ \cite{cartarius:2017}:
\begin{eqnarray}
\label{eq:compressibility}
		\Delta n_j = \langle \hat n_j^2\rangle - \langle \hat n_j\rangle^2 \,.
\end{eqnarray} 
This quantity is connected to the local compressibility \cite{Local:Compressibility}. A phase is classified as incompressible when $\Delta n_j$ vanishes at all sites $j$. 
Let us note that while measurement of $\Delta n_j$ requires single site resolution other possible methods allow to access compressibility in harmonic traps \cite{Roscilde2009,Delande2009}.
 
The superfluid phase is signalled by the non-vanishing value of single particle correlations $\langle \hat{a}_i^\dagger \hat{a}_j\rangle$ across the lattice. In particular, we analyze the Fourier transform of the off-diagonal single particle correlations, which is defined as \cite{Jiang2012pair}:
\begin{equation}
M_1(q)= \frac{1}{L^2} \sum_{i,j=1}^{L-1} e^{i q\left( i - j\right)} Re\langle \hat{a}_i^\dagger \hat{a}_j\rangle\label{bcd}
\end{equation}
The phase is a superfluid when $M_1(q)\neq 0$ and the maximum Fourier component is $q=q_{\rm max}$ with $q_{\rm max}=0$. When  the maximum of $M_1(q)$ is at $q_{\rm max}=\pi$ the phase is a staggered superfluid (SSF) \cite{Johnstone19}.

Diagonal long-range order is revealed by a peak of the static structure form factor $S(q)$ at the corresponding Fourier component, where
\begin{equation}
S(q)=\frac{1}{L^2} \sum_{i,j=1}^{L-1} \left\langle \hat{n}_i\hat{n}_j \right\rangle e^{-iq\left(j-i\right)} \label{DW_orderparameter} \ . 
\end{equation}
A single peak of $S(q)$ at $q=2\pi/j$ signals a periodic structure with periodicity $ja$. We denote this phase by charge density wave CDW$_j$ if it is incompressible. If instead the phase is superfluid, it is denoted by lattice supersolid phase $j$. We distinguish between two kinds of lattice supersolid phases, depending on the Fourier spectrum of the single-particle off-diagonal correlations. The phase is a lattice supersolid SS$_j$ when $q_{\rm max}=0$. If the peak is instead at $q_{\rm max}=\pi$, then the phase is a staggered supersolid  SSS$_j$ \cite{Johnstone19}.

Additionally, pair tunneling terms are expected to favour the onset of what has been denoted by pair superfluidity (PSF)\cite{sowinski2012dipol,Luehmann2016,Juergensen2015,Johnstone19,biedron2018extended,Dutta2011}. For the parameter regime of our study we do not find PSF, but for completeness we report the observables we use in order to come to our conclusions. PSF is signalled by a non-vanishing expectation value of the pair-correlation function. In this work we analyze the Fourier transform of the pair correlations $\left\langle \hat{a}^\dagger_{i} \hat{a}^\dagger_{i} \hat{a}_{j}\hat{a}_{j} \right\rangle$, which we define as \cite{Jiang2012pair}:
\begin{equation}
M_2(q) = \frac{1}{L^2}\sum_{i,j=1}^{L-1} e^{i q\left(i-j \right)} {\rm }Re\left\langle \hat{a}_i^\dagger \hat{a}_{i}^\dagger \hat{a}_{j} \hat{a}_{j}\right\rangle \ . \label{pcd}
\end{equation}
In the pair superfluid (PSF) and in the pair supersolid (PSS) phases the Fourier components of $M_2(q)$, Eq. (\ref{pcd}), are larger than the corresponding Fourier components of $M_1(q)$, Eq. (\ref{bcd}) \cite{Jiang2012pair}. 

The MI phase is characterized by vanishing off-diagonal correlations, vanishing compressibility, and $S(q)=0$. We verify the existence of the Haldane insulator by calculating the expectation value of the modified string-order parameter  \cite{Batrouni2013,Qin:2003}
\begin{equation}
O_s(r)=\langle \delta \hat{n}_i e^{ \left(i\theta \sum_{k=i}^{i+r}\delta \hat{n}_k \right)} \delta \hat{n}_{i+r}\rangle
\label{eq:string}
\end{equation}
with $\delta \hat{n}_i = \hat{n}_i-\rho$. For density $\rho=2$ we take $\theta=\pi/2$ \cite{Qin:2003}. 
Finally, we analyze the entanglement entropy for a partition of the chain into two equally long subsystems A and B: 
$$S_{\text{vN}}=-{\rm Tr}\{\hat \rho_B\ln\hat \rho_B\}\,,$$ 
where $\hat \rho_B={\rm Tr}_A\{|\phi_0\rangle\langle \phi_0|\}$ and $|\phi_0\rangle$ is the ground state. We refer the readers to Appendix B for further details on the numerical implementations (including how we treat the boundary effects). Table \ref{Table:1} summarizes the expectation values that characterize the phases here discussed. 

\begin{widetext}
	
	\begin{table}
		\begin{tabular}{|p{3cm}|p{1.8cm}||p{2.3cm}|p{2.3cm}|p{2.3cm}|p{2.3cm}|p{2cm}| } 
			\hline
			Phase & Acronym & $\Delta n_i$, Eq. \eqref{eq:compressibility} & $M_1(q)$, Eq. \eqref{bcd} & $M_2(q)$, Eq. \eqref{pcd} & $S(q)$, Eq. \eqref{DW_orderparameter} &$O_s$, Eq. \eqref{eq:string} \\ [3ex]
			\hline\hline
			Mott-Insulator & MI   & $0$    &$0$&   $0$&$q_{\rm max}=0$ & 0 \\
			Density Wave & CDW$_j$&   $0$  & $0$   &$0$& $q_{\rm max}=2\pi/j$ & $\neq 0$ \\
			Haldane-Insulator & HI   & $0$    &$0$&   $0$&$q_{\rm max}=0$ & $\neq 0$ \\
			Superfluid & SF & $\ne 0$ & $q_{\rm max}=0$&  $M_2(q)<M_1(q)$& $q_{\rm max}=0$ & 0\\
			Staggered Superfluid & SSF&   $\ne 0$  & $q_{\rm max}=\pi$&$M_2(q)<M_1(q)$& $q_{\rm max}=0$ & 0\\
			Supersolid & SS$_j$    &$\ne 0$ & $q_{\rm max}=0$&  $M_2(q)<M_1(q)$& $q_{\rm max}=2\pi/j$ & 0 \\
			Staggered Supersolid & SSS$_j$& $\ne 0$  & $q_{\rm max}=\pi$   &$M_2(q)<M_1(q)$& $q_{\rm max}=2\pi/j$ & 0\\
			Pair Superfluid & PSF & $\ne 0$  &$M_1(q)<M_2(q)$& $q_{\rm max}=0$& $q_{\rm max}=0$ & 0\\
			Pair Supersolid & PSS$_j$ & $\ne 0$& $M_1(q)<M_2(q)$& $q_{\rm max}=0$& $q_{\rm max}=2\pi/j$ & 0 \\
			\hline
		\end{tabular}
		\caption{Table of the phases, of their acronyms, and of the corresponding values of the observables.	The subscript $j$ of the Density Wave and of the Supersolid phases refer to the component $q=2\pi/j$ of the structure form factor which is different from zero, correspondingly the density modulation has periodicity $ja$. $q_{\rm max}$ indicates the Fourier component at which the spectra of $M_1$, $M_2$, $S(q)$ may have a maximum. }\label{Table:1}

	\end{table}
	
\end{widetext}

\subsection{Nearest-Neighbour interactions}
\label{Sec:nn}

\begin{figure}[h!]
	\centering
		\includegraphics[width=0.5 \textwidth]{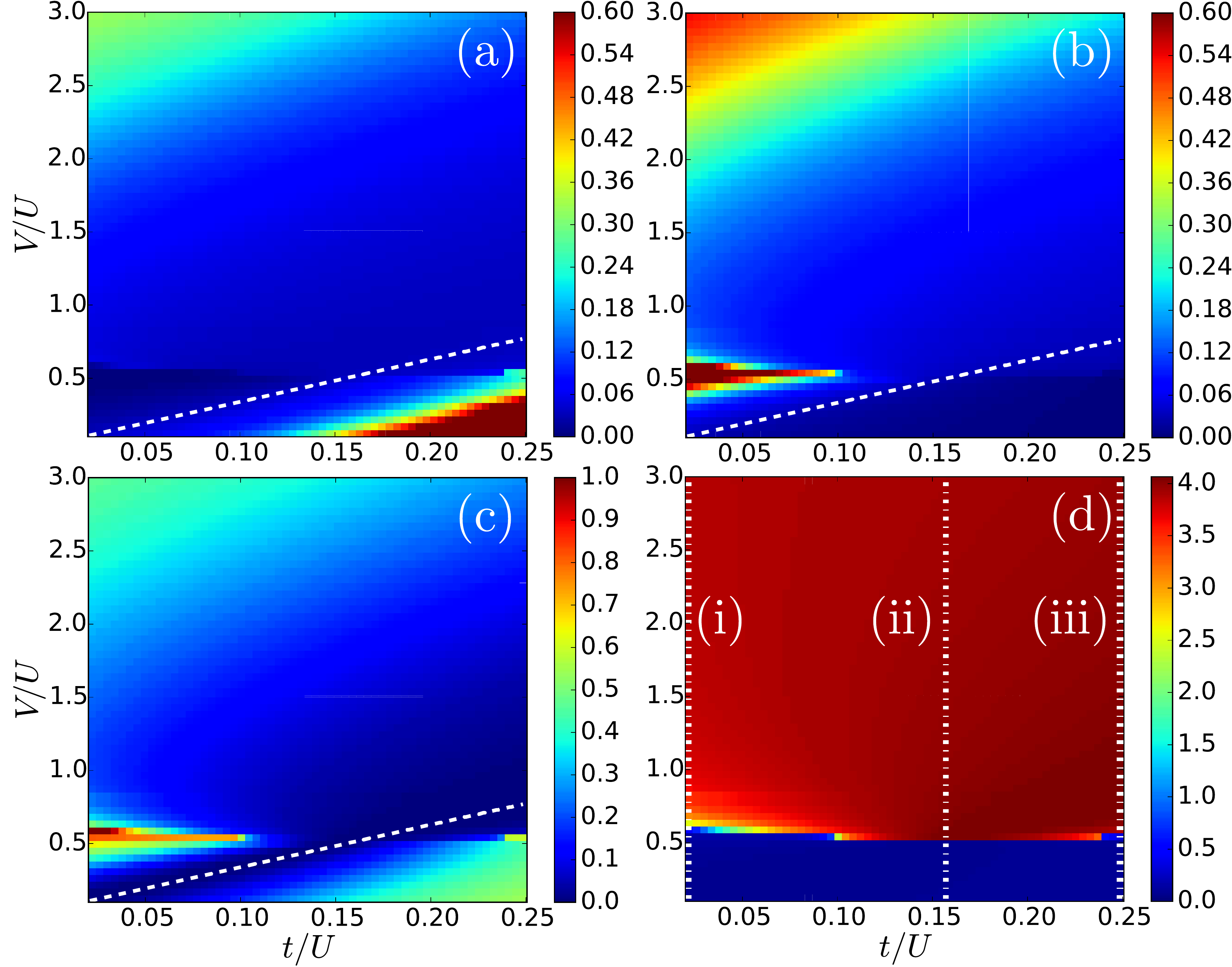}
\caption{(color online) Contour plot of the relevant observables in the $V/U-t/U$ plane for the ground state of the Bose-Hubbard model of Eq. \eqref{HBH:nn}. Subplot (a) and (b) signal SSF and SF through the Fourier components $q=0$ and $q=\pi$, respectively, of the single particle off-diagonal correlations $M_1(q)$. Subplot (c) reports the maximum value of the compressibility across the lattice, Eq. \eqref{eq:compressibility}, and subplot (d) the component at momentum $\pi$ of the structure form factor, signaling the onset of a density modulation. The number of lattice sites is fixed to $L=60$ and the number of particles is given by $N=120$. The white dashed line in (a)-(c) indicates the interaction-induced atomic limit. The vertical dotted lines in (d) indicate the parameters of the sweeps in Fig. \ref{fig:Struct:cut:nn}. See Appendix \ref{App:DMRG} for further details. } 
	\label{fig:PD:nn}
\end{figure}

The properties of the ground state of the Bose-Hubbard Hamiltonian $H_{\rm BH}^{(1)}$, Eq. \eqref{HBH:nn}, are studied for a finite chain at density $\rho=2$ and as a function of the dipolar interaction strength $d$, Eq. \eqref{Eq:d}. We report the phase diagrams as a function of the  blockade coefficient between nearest-neighbor, $V/U$, and of the tunneling rate $t/U$. Figures \ref{fig:PD:nn}(a)-(d) display the contour plots of the relevant observables for (a) the SF phase, (b) the SSF phase, (c) the incompressible phase and (d) the phases with diagonal long range orders. 

We first identify the MI-SF phase transition at $V\to 0$: the transition point $(t/U)_c$ is in qualitative agreement with the literature \cite{Knap:2012}, the discrepancy is attributed to the finite size of the chain.
At finite and nonvanishing ratios $V/U$ the incompressible phase moves to larger values of $t/U$. It is localized about the white dashed line, which indicates the atomic limit due to quantum interference (see Sec. \ref{Sec:3}). By inspecting subplots (a) and (b) it is evident that this phase divides the diagram into two disconnected SF phases: According to our classification, on the left the phase is a staggered SF (SSF), on the right it is a SF. It is remarkable that also at $t/U\to 0$ the phase is superfluid. According to our preliminary considerations, this superfluid phase is due to the correlated hopping of the dipolar interactions.

Inspecting the single-particle off-diagonal correlations, subplot (a), and the structure form factor, subplot (d), we further observe a transition about the line $V=U/2$. The properties at this transition depend on $t/U$. We recall that at this value and in the atomic limit we expect a first-order transition from a MI to a CDW[4,0] \cite{Batrouni2006}. This is consistent with our numerical results along the values of the interaction-induced atomic limit. When hopping is dominated by the kinetic energy (on the right of the interaction-induced atomic limit) the phase is expected to undergo a continuous transition from SF to SS (not included in our phase diagram). When instead hopping is due to interactions we observe a continuous transition from SSF to staggered SS (SSS). 

In the next sections we discuss some of these behaviors in detail.

\subsubsection{Superfluidity}

\begin{figure}[h!]
	\centering
	\begin{minipage}[t]{0.45\textwidth}\vspace{0pt}
		\includegraphics[width=\textwidth]{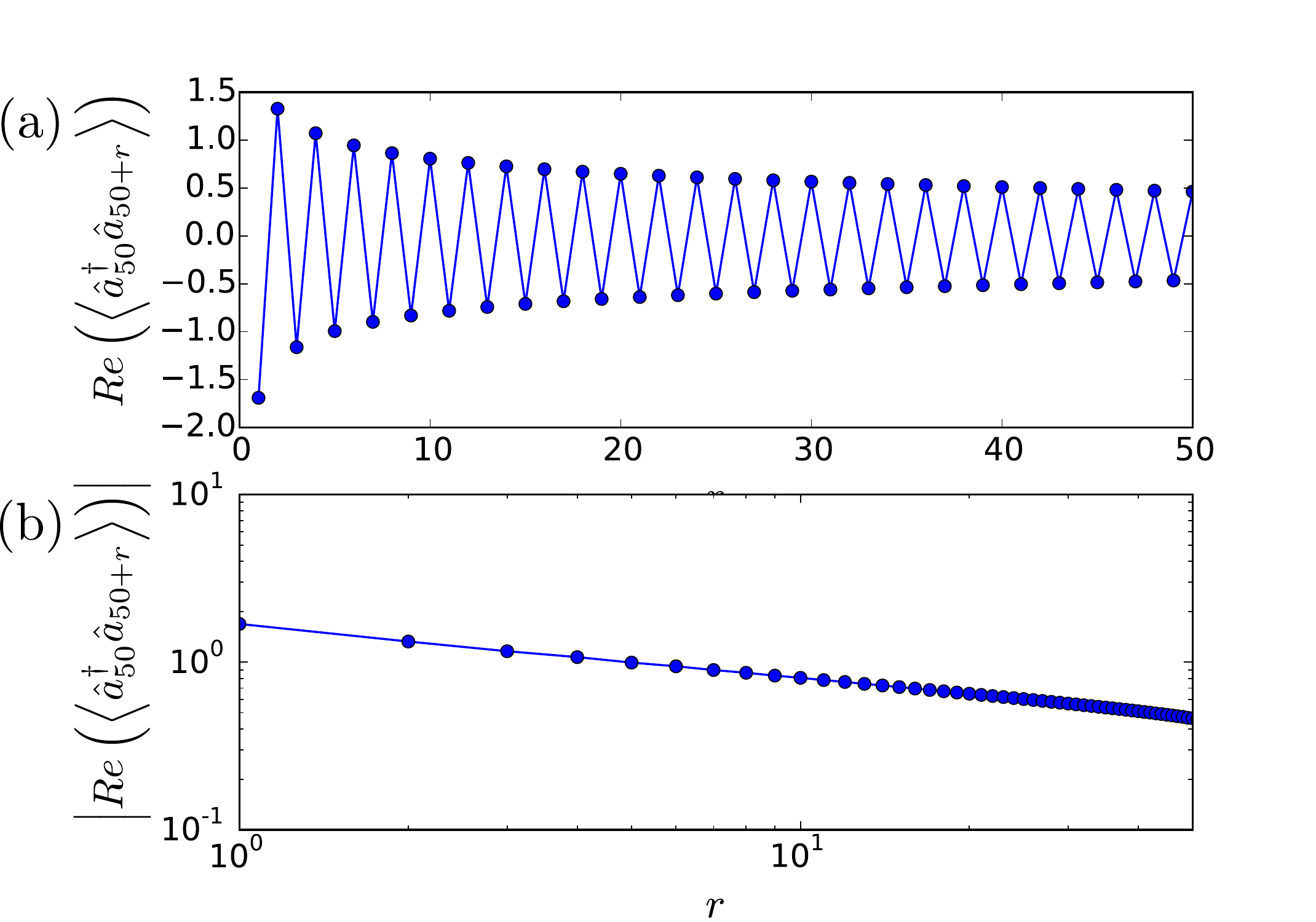}
	\end{minipage}\hfill%
	\caption{(color online) (a) Single particle off-diagonal correlation $\left\langle\hat{a}^\dagger_{\ell}\hat{a}_{\ell+r}\right\rangle$ for a particle at the center of the chain and as a function of the distance $r$ (in units of the lattice constant $a$). Here, $V/U=0.529$ and $t/U=0.042$, corresponding  to a point in the SSF phase.  The chain has $L=120$ sites and $N=240$ particles, the site close to the center is $\ell=50$. (b) Same as (a) but in logarithmic scale. Here the absolute value of the correlation function is reported and the power-law decay is evident.} 
	\label{fig:SF}
\end{figure}

\begin{figure}[h!]
	\centering
	(a)
	\begin{minipage}[t]{0.4\textwidth}\vspace{0pt}
		\includegraphics[width=\textwidth]{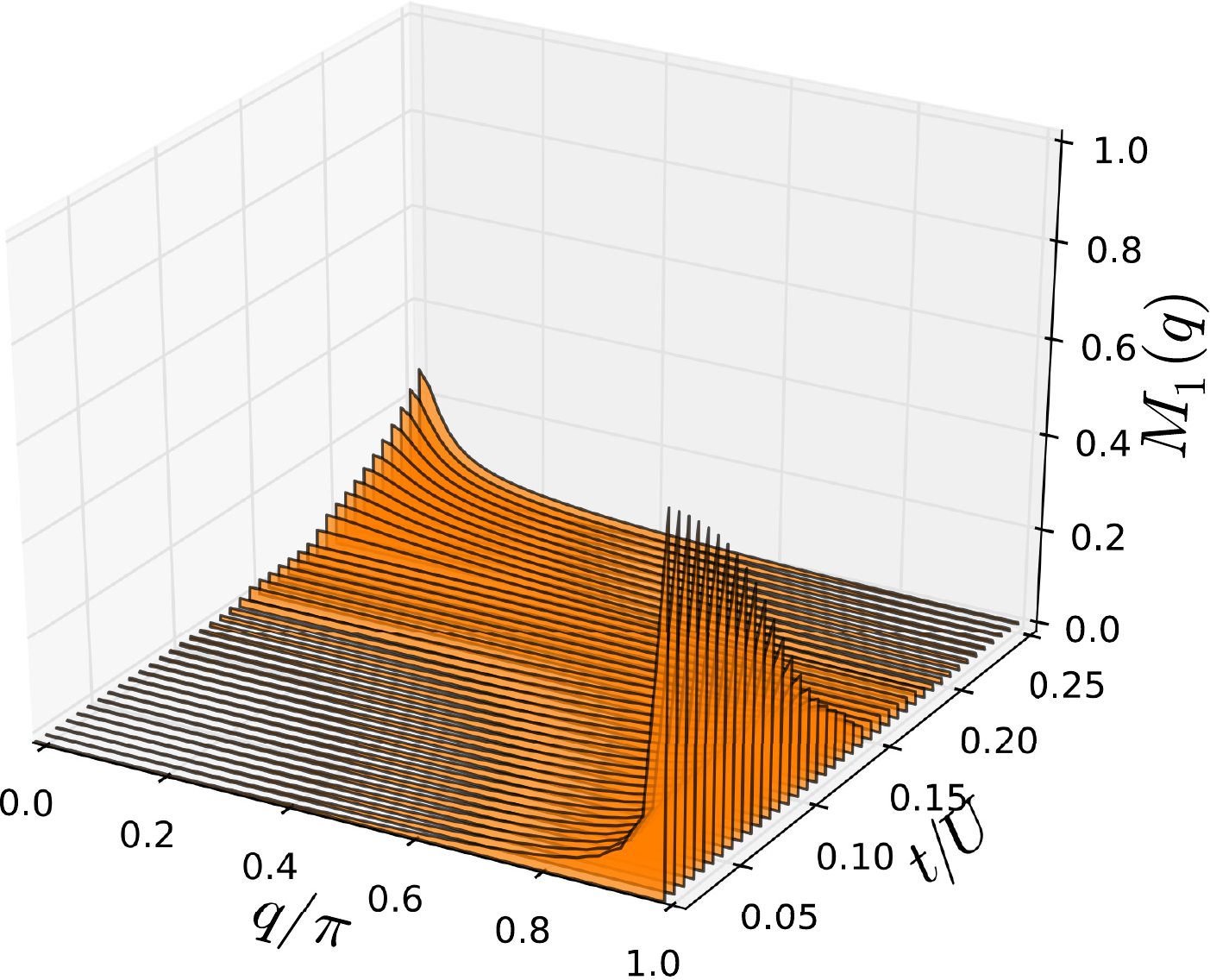}
	\end{minipage}\hfill\\
	(b)
	\begin{minipage}[t]{0.45\textwidth}\vspace{0pt}
		\includegraphics[width=\textwidth]{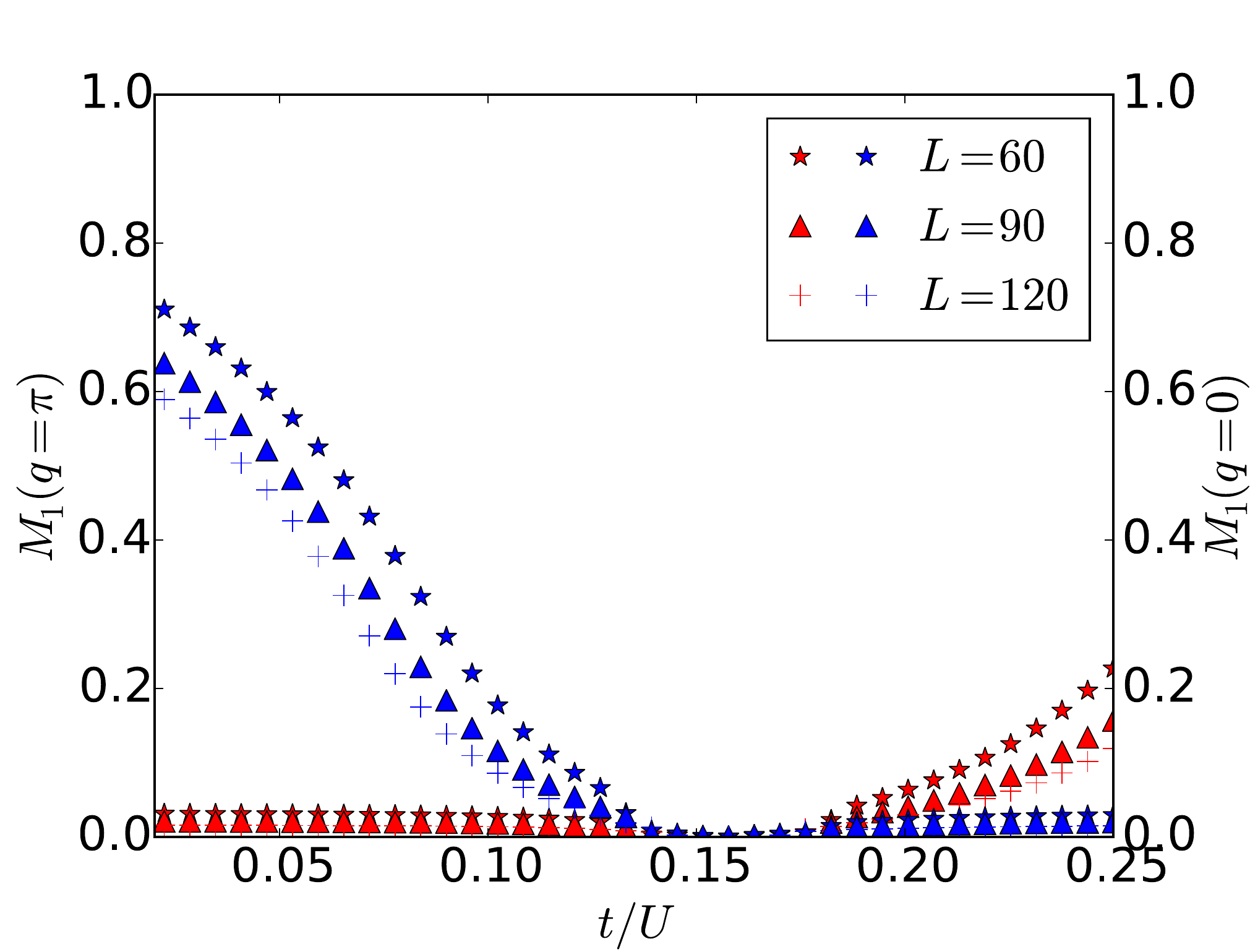}
	\end{minipage}\hfill\\
	(c)
	\begin{minipage}[t]{0.45\textwidth}\vspace{0pt}
		\includegraphics[width=\textwidth]{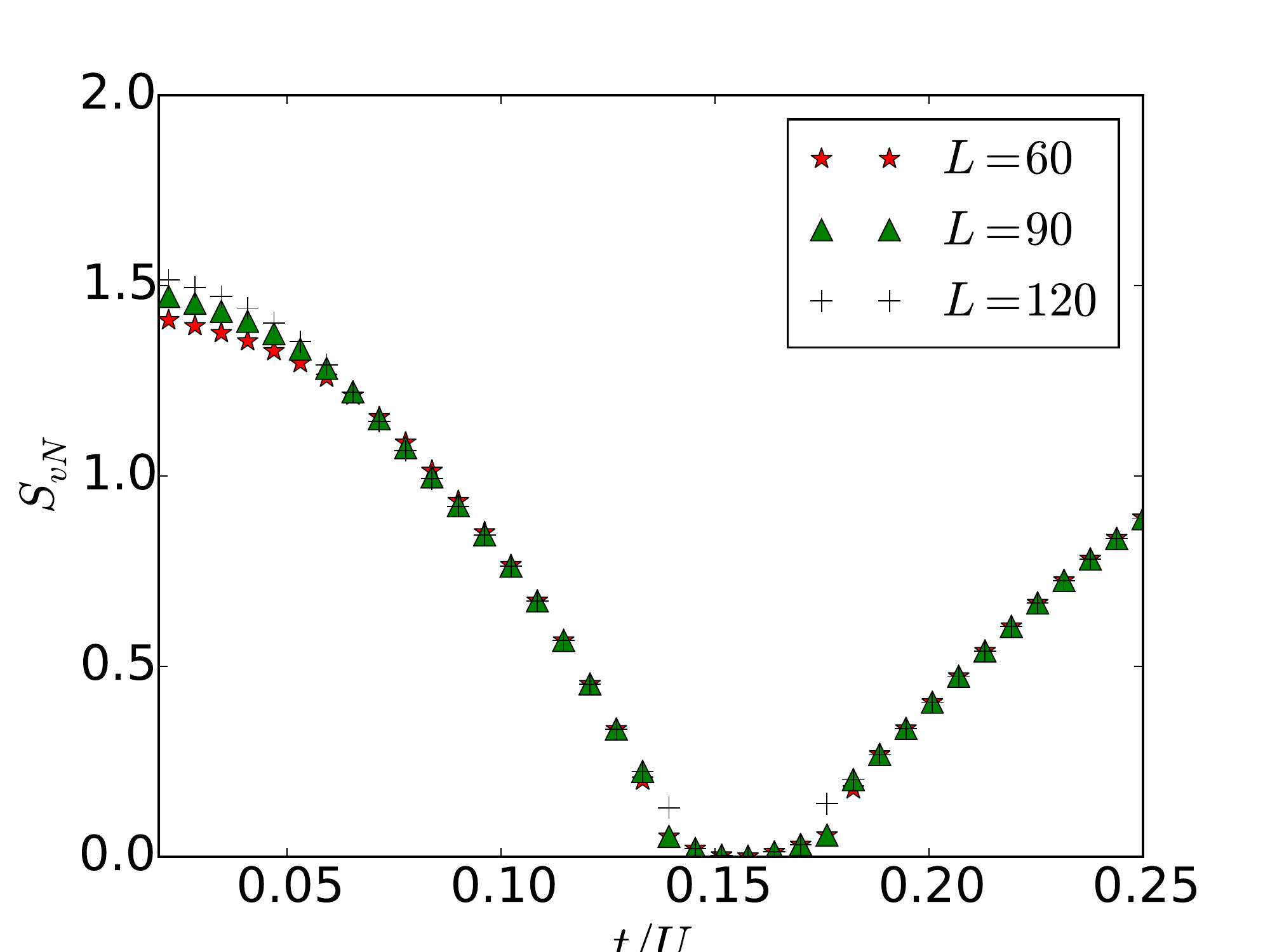}
	\end{minipage}\hfill\\
	\caption{(color online) (a) Fourier transform of the single particle off-diagonal correlations $M_1(q)$ (\ref{bcd}) as a function of $q$ and of $t/U$ for $V/U=0.5$. Subplot (b) shows the behavior of the Fourier components at $q=0$ (red) and $q=\pi$ (blue) as a function of $q$. Subplot (c) displays corresponding values of the entanglement entropy. Different symbols correspond to different system sizes $L$ ($L=60,90,120$, see legenda), keeping $N=2L$. Subplot (a) is reported for $L=60$.}
	\label{fig:5}
\end{figure}

\begin{figure}[h!]
	\centering
	(a)
	\begin{minipage}[t]{0.4\textwidth}\vspace{0pt}
		\includegraphics[width=\textwidth]{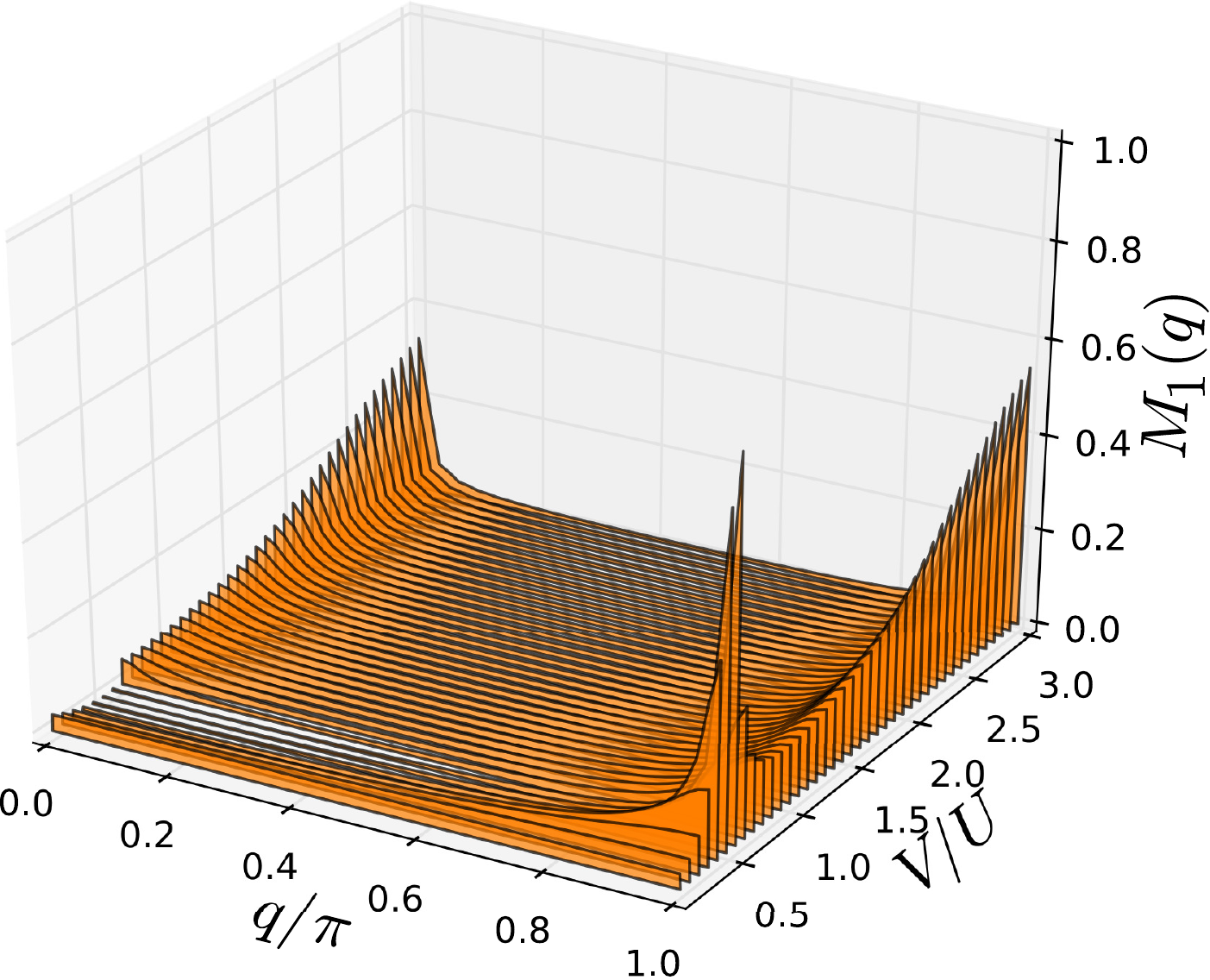}
	\end{minipage}\hfill\\
	(b)
	\begin{minipage}[t]{0.45\textwidth}\vspace{0pt}
		\includegraphics[width=\textwidth]{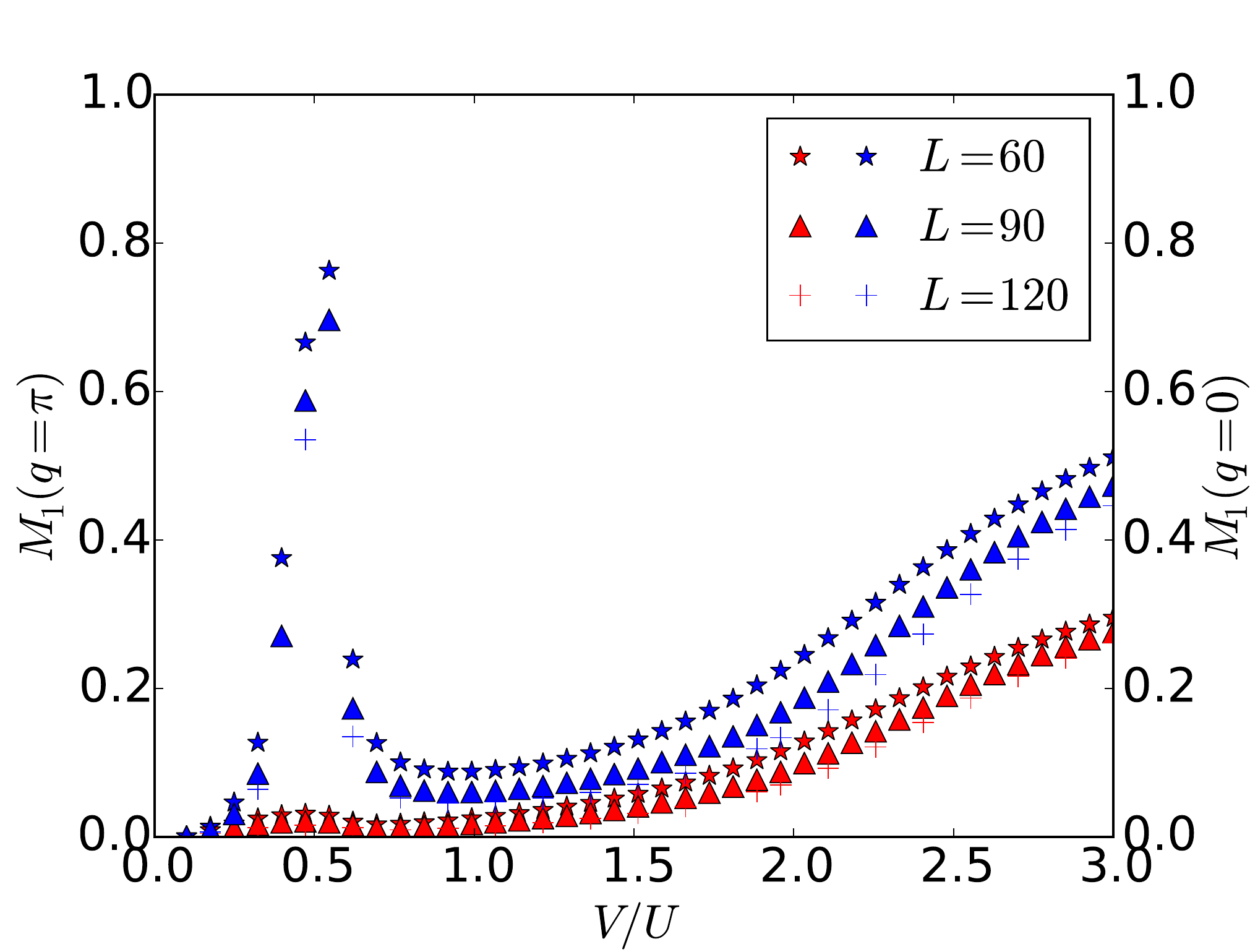}
	\end{minipage}\hfill\\
	(c)
	\begin{minipage}[t]{0.45\textwidth}\vspace{0pt}
\includegraphics[width=\textwidth]{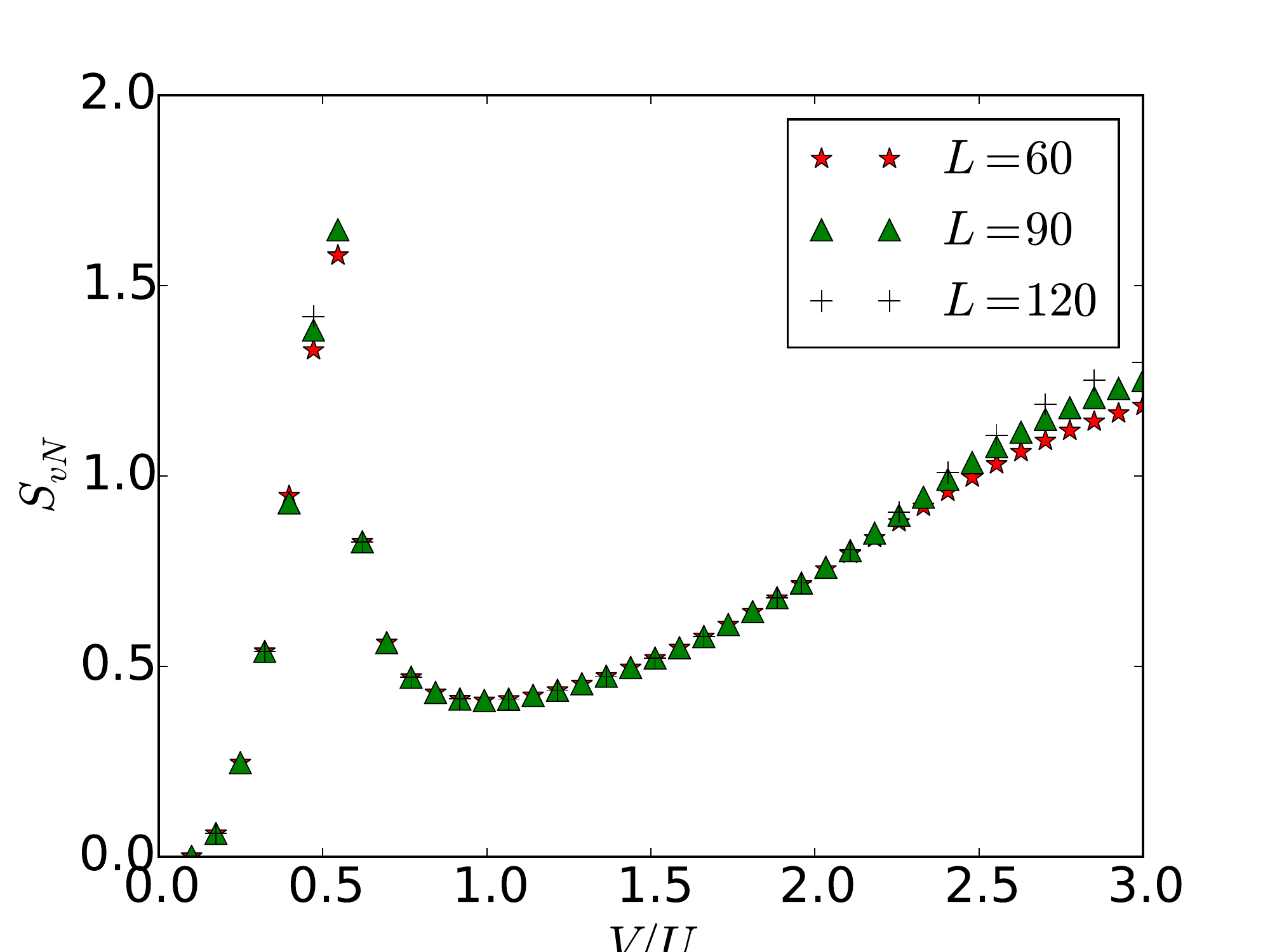}
	\end{minipage}\hfill\\	
	\caption{(color online) Same as Fig. \ref{fig:5} but as a function $V/U$ and taking $t=0.02U$. The peak is located in the SSF region visible in Fig. \ref{fig:SF}(c) at $V/U=0.5$. The behavior of the entanglement entropy suggests a continuous phase transition at this point. } 
	\label{fig:6}
\end{figure}

In one-dimension there is no long-range off-diagonal order, and superfluidity is signalled by the power-law decay of the single-particle correlation function with the distance \cite{Giamarchi2004}. In the SSF phase this behavior is modulated by an oscillation with (dimensionless) wave number $q=\pi$, so that the correlation function changes sign every time the distance is increased by one lattice site. This oscillation is visible in Fig. \ref{fig:SF}(a), which reports the off-diagonal correlation in the parameter regime of the SSF phase. Subplot (b) provides evidence of the power-law decay of the envelope.  We note that this behavior was also reported in Refs. \cite{sowinski2012dipol,biedron2018extended}, and was there denoted by "pair superfluidity". We consider here more appropriate to denote this phase by "staggered SF", since it is due to the dominant contribution of the density-assisted tunneling term in establishing off-diagonal correlations. Interestingly, the oscillation is already captured by a mean-field model, cf. Sec. \ref{Sec:3}. In general, the analysis of the Fourier transform of $M_1(q)$ across the diagram shows that in the superfluid phase the Fourier components different from zero are solely at $q=0$ and $q=\pi$. This is visible, for instance, in Figs. \ref{fig:5}(a) and \ref{fig:6}(a). 

.

We now analyze the onset and the features of the SF phase along two specific transition lines: along the axis $V/U$ for $t/U=0.02$ and along the axis $t/U$ for $V=U/2$. 

We first consider $V=U/2$. Figure \ref{fig:5} (a) displays the Fourier spectrum of the single particle correlation function, $M_1(q)$, as a function of $t/U$: The non-vanishing Fourier components are at $q=0$ (at large $t/U$) and at $q=\pi$ (at small $t/U$). These two Fourier components are reported in subplot (b): they are different from zero on the right and on the left, respectively, of the interaction-induced atomic limit. When moving towards the atomic limit they decrease until they vanish.  The entanglement entropy (c) vanishes for an interval of values centered about the transition point.

The behavior of superfluidity for small ratios $t/U$ is determined by the correlated hopping. Figure \ref{fig:6}(a) displays $M_1(q)$ as a function of $V/U$ and small ratio $t/U$. In subplot (b) we report the Fourier components at $q=0$ and $q=\pi$. The phase is incompressible for a small interval about $V=0$, after which the Fourier component at $q=\pi$ rapidly grows and reaches a maximum about $V=U/2$. 
At the same value the entanglement entropy, (c) displays a maximum. After this maximum, the Fourier component at $q=\pi$ drops to smaller values, while the $q=0$ component starts to grow from zero to a small but finite value (the numerics converge very slowly at these points and we cannot provide more detailed sampling). The Fourier component at $q=\pi$ is always larger than $M_1(0)$, therefore according to our definition the SF phase is staggered.  We have analyzed the scaling of the peak at $V=U/2$ with the system size: by means of a fit we extract that the peak height at the asymptotics is finite and tends to the finite value $M_1(\pi)\to 0.47$ (see Appendix C for further details). 

We note that across the phase diagram we do not observe pair superfluidity: In fact, the expectation value of observable $M_2$ is always smaller than the one of $M_1$ at the corresponding $q$ value (see Appendix \ref{App:PD:Details} for details). Moreover, we do not find spatial modulation in the pair correlations. We believe that this is because the pair tunneling coefficient $P$ is negative over the considered parameter range. 

\subsubsection{Diagonal long-range order}

\begin{figure}[h!]
	\centering
	\begin{minipage}[t]{0.45\textwidth}\vspace{0pt}
		\includegraphics[width=\textwidth]{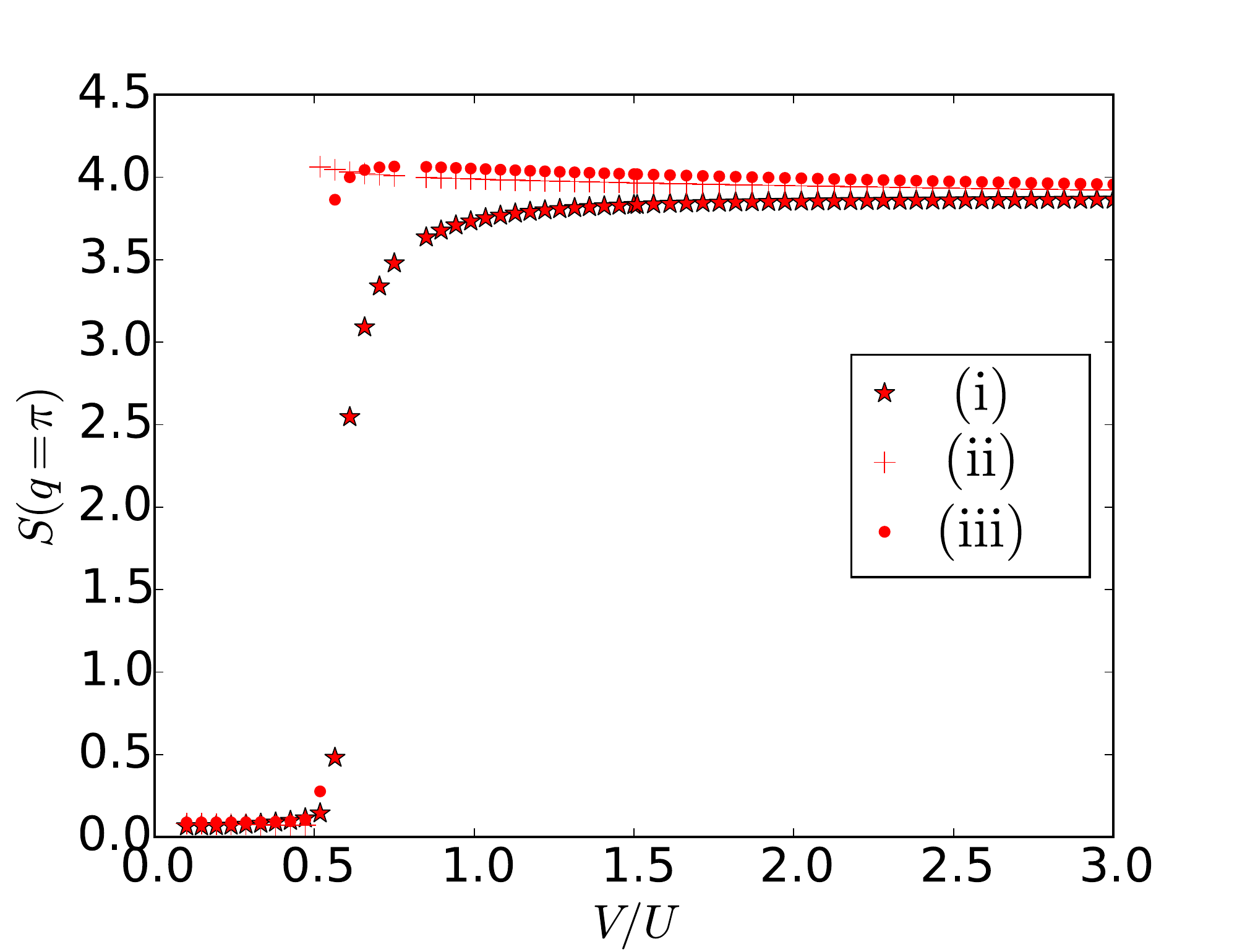}
	\end{minipage}\hfill\\
	\caption{(color online) The component at $q=\pi$ of the structure form factor, Eq. \eqref{DW_orderparameter}, signaling the onset of density modulations. The data are taken at the sweeps (i), (ii), (iii) of Fig. \ref{fig:PD:nn}(d). Here, $t/U=0.02$ at (i), $t/U=0.157$ at (ii), and $t/U=0.25$ at (iii).} 
\label{fig:Struct:cut:nn}
\end{figure}

Let us now discuss the onset of long-range order. This is here signalled by the non-vanishing component at $q=\pi$ of the structure form factor. Figure \ref{fig:Struct:cut:nn} displays its behavior as a function of $V/U$ for three different values of $t/U$. The parameters of these sweeps are indicated by the vertical lines in Fig. \ref{fig:PD:nn}(d). The data of (i) correspond to the sweep across the transition from SSF to SSS and show a continuous, even though rapid, growth of the structure form factor. This rapid growth occurs in the same parameter interval where the superfluid Fourier component at $q=0$ increases from zero to a finite value. 

Sweep (ii) is taken across the transition MI-CDW[4,0]. It shows a discontinuity at $V/U=0.5$, indicating a first-order phase transition. This agrees with the mean-field prediction. Sweep (iii) moves across the SF to the incompressible CDW[4,0] phase. The behavior suggests a discontinuous, first-order transition. We also expect a continuous transition SF to SS at $V=U/2$ but for slightly larger ratios $t/U$, that are not included in this phase diagram.

We have also calculated the string-order parameter given by Eq. (\ref{eq:string}) across the phase diagram. We could not identify a Haldane Insulator phase (see Appendix \ref{App:PD:Details}). This result is consistent with the literature. In fact, Monte-Carlo simulations could not find the Haldane insulator for $\rho=2$ in the extended Bose-Hubbard model without correlated hopping terms. Moreover, for density $\rho=1$ correlated hopping tends to shrink the Haldane phase \cite{biedron2018extended}. 

\begin{figure}[h!]
	\centering
	\begin{minipage}[t]{0.45\textwidth}\vspace{0pt}
		\includegraphics[width=\textwidth]{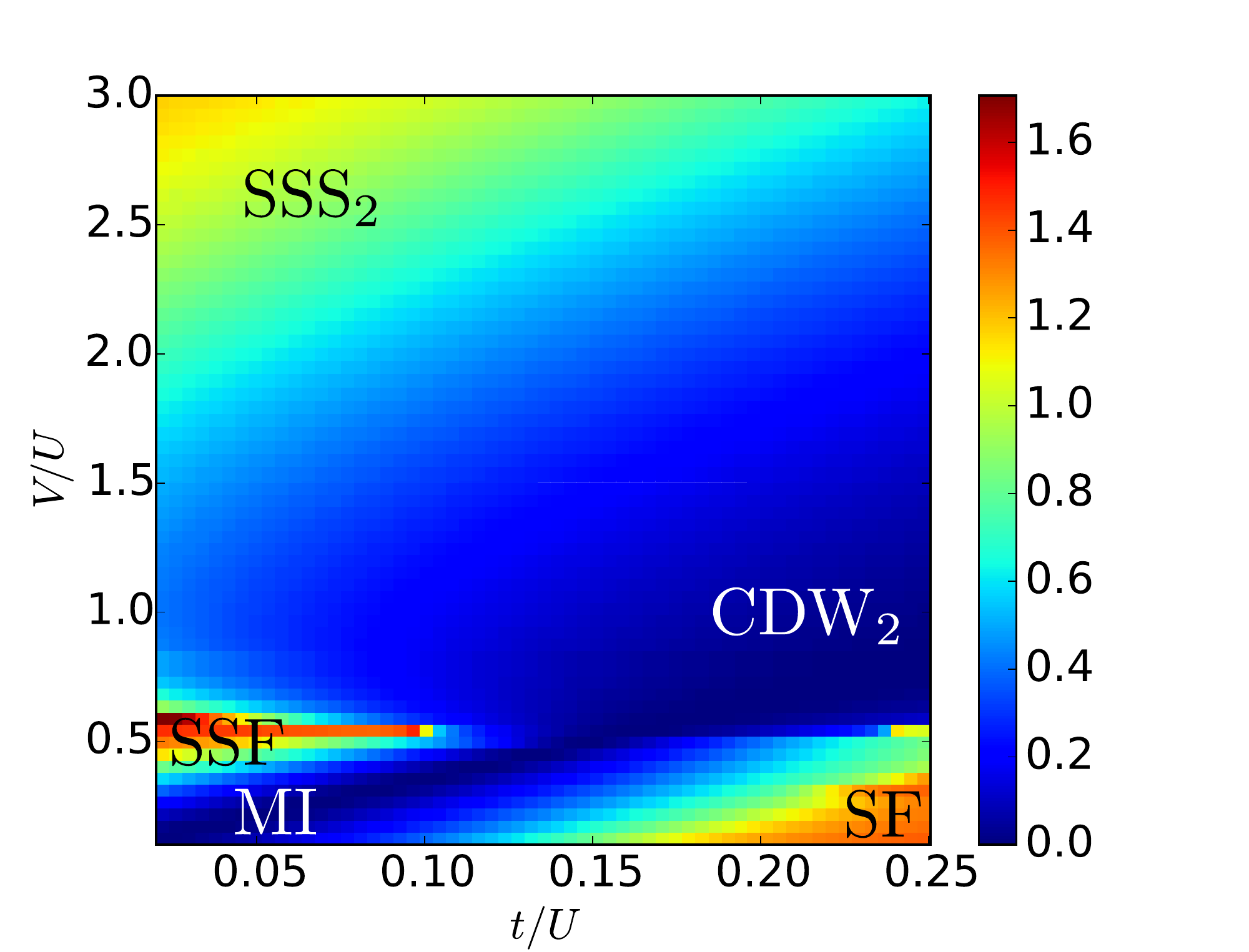}
	\end{minipage}\hfill\\
	\caption{(color online) Contour plot of the von Neumann entropy for the ground state of Eq. \eqref{HBH:nn} as a function of the ratio $V/U$ and of the ratio $t/U$. The phases are labeled according to the classification of Table \ref{Table:1}. The parameters are the same as in Fig. \ref{fig:PD:nn}, the size of the subsystem is $L_B=30$ sites. } 
	\label{fig:7}
\end{figure}

Figure \ref{fig:7} displays the contour plot of the entanglement entropy. The region with non-vanishing values are superfluid phases. We observe in particular the maximum at the transition from SSF to SSS at $t/U\simeq 0$ and $V\simeq U/2$. We label the phases in the diagram according to our classification in Table \ref{Table:1}.

\subsection{Next-nearest-neighbors interactions}
\label{Sec:nnn}

\begin{figure}[h!]
	\centering
		\includegraphics[width=0.5\textwidth]{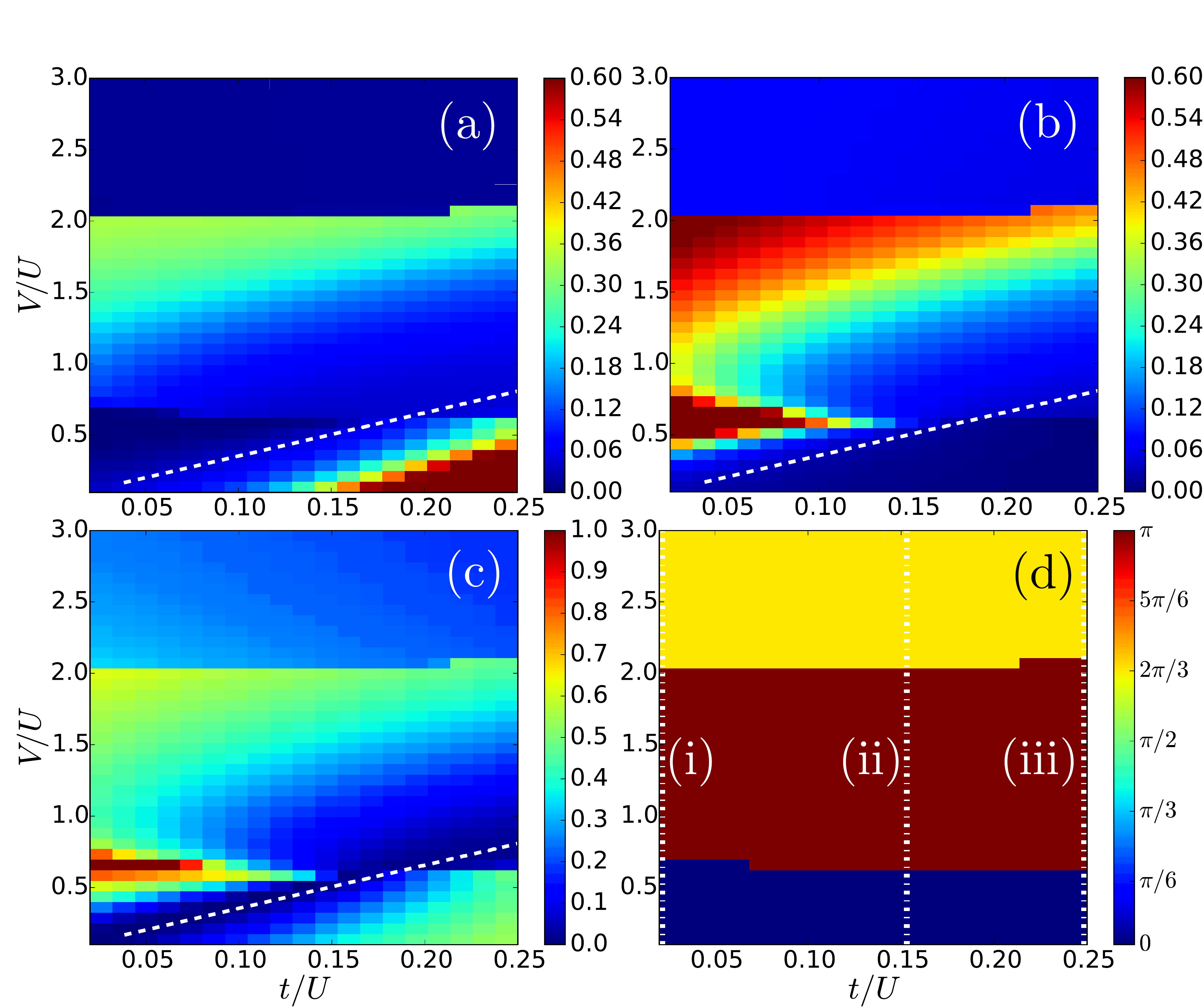}	
\caption{(color online) Contour plot of relevant observables in the $V/U-t/U$ plane for the ground state of the Bose-Hubbard model of Eq. \eqref{HBH:nnn}, which includes next-nearest neighbor interactions. Subplot (a) and (b) signal SSF and SF through the Fourier components $q=0$ and $q=\pi$, respectively, of the single particle off-diagonal correlations $M_1(q)$. Subplot (c) reports the maximum value of the compressibility across the lattice, Eq. \eqref{eq:compressibility}, and subplot (d) the component at momentum $\pi$(red) and $2\pi/3$ (yellow) of the structure form factor, signalling the onset of a density modulation with the corresponding periodicity. The number of lattice sites is fixed to $L=60$ and the number of particles is given by $N=120$. The white dashed line in (a)-(c) indicates the interaction-induced atomic limit. The vertical dotted lines in (d) indicate the parameters of the sweeps in Fig. \ref{fig:Struct:cut:nnn}. See Appendix \ref{App:DMRG} for further details. } 
	\label{fig:PD:nnn}
\end{figure}

We now analyze the ground state of the Bose-Hubbard Hamiltonian with next-nearest-neighbors interactions. The Hamiltonian is given in Eq. \eqref{HBH:nnn}. The properties of the relevant observables are shown in Fig. \ref{fig:PD:nnn}. They share some similarities with the nearest-neighbor model, (compare with Fig. \ref{fig:PD:nn}). For instance, also in this case we observe an incompressible phase at the interaction-induced atomic limit, which separates staggered superfluidity from "normal" superfluidity. However, now the SF phases occur in larger parameter regions and the incompressible phase shrinks. 
Moreover, the transition to the diagonal long-range order is located about $V\sim 0.5 U$, even though it is shifted to a slightly larger value than for the nearest-neighbor case. A striking difference is the appearance of a third phase at $V\sim 2U$, which is signalled by a peak of the structure form factor at $q=2\pi/3$ . 

\begin{figure}[h!]
	\centering
	(a)
	\begin{minipage}[t]{0.45\textwidth}\vspace{0pt}
		\includegraphics[width=\textwidth]{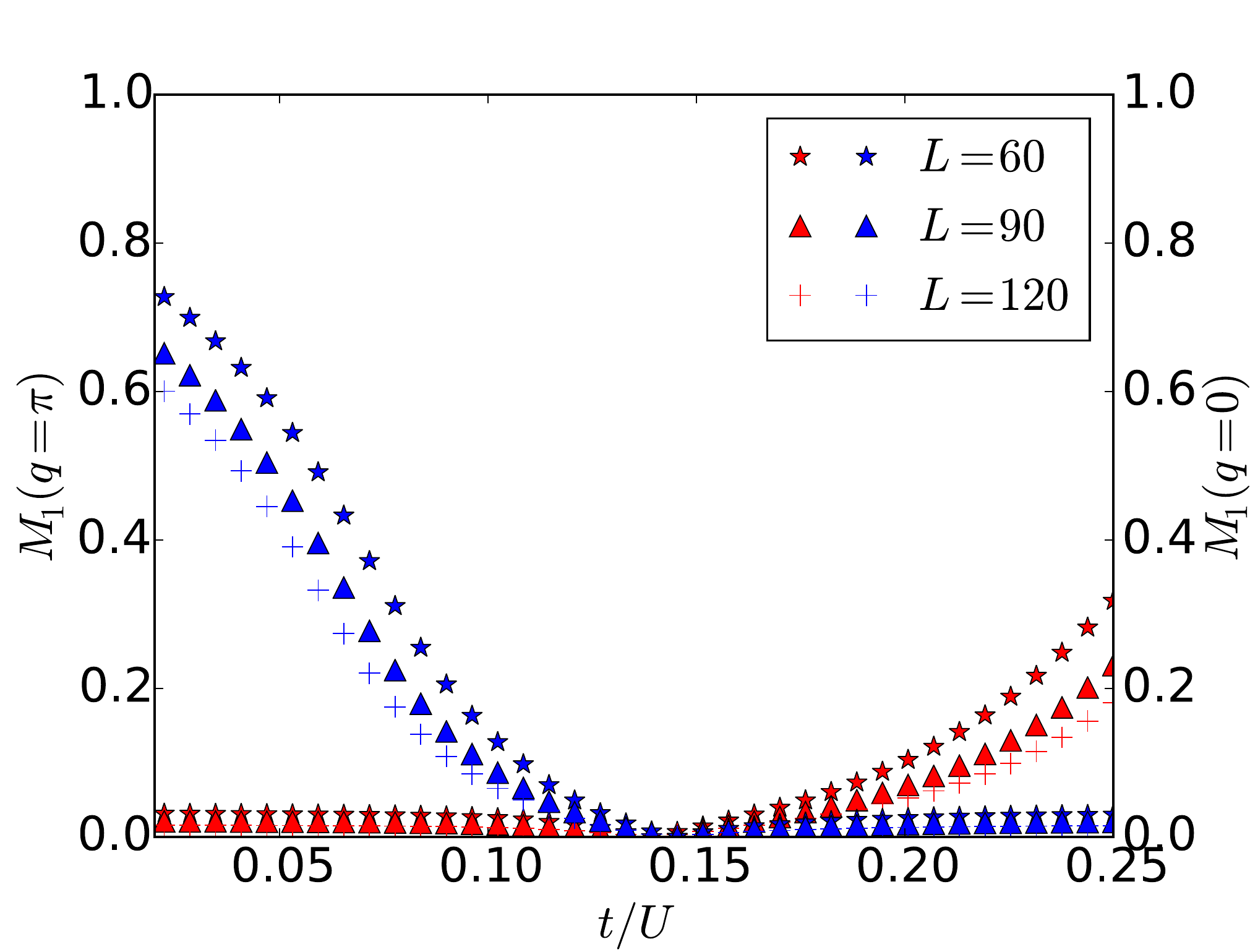}
	\end{minipage}\hfill\\
	(b)
	\begin{minipage}[t]{0.45\textwidth}\vspace{0pt}
		\includegraphics[width=\textwidth]{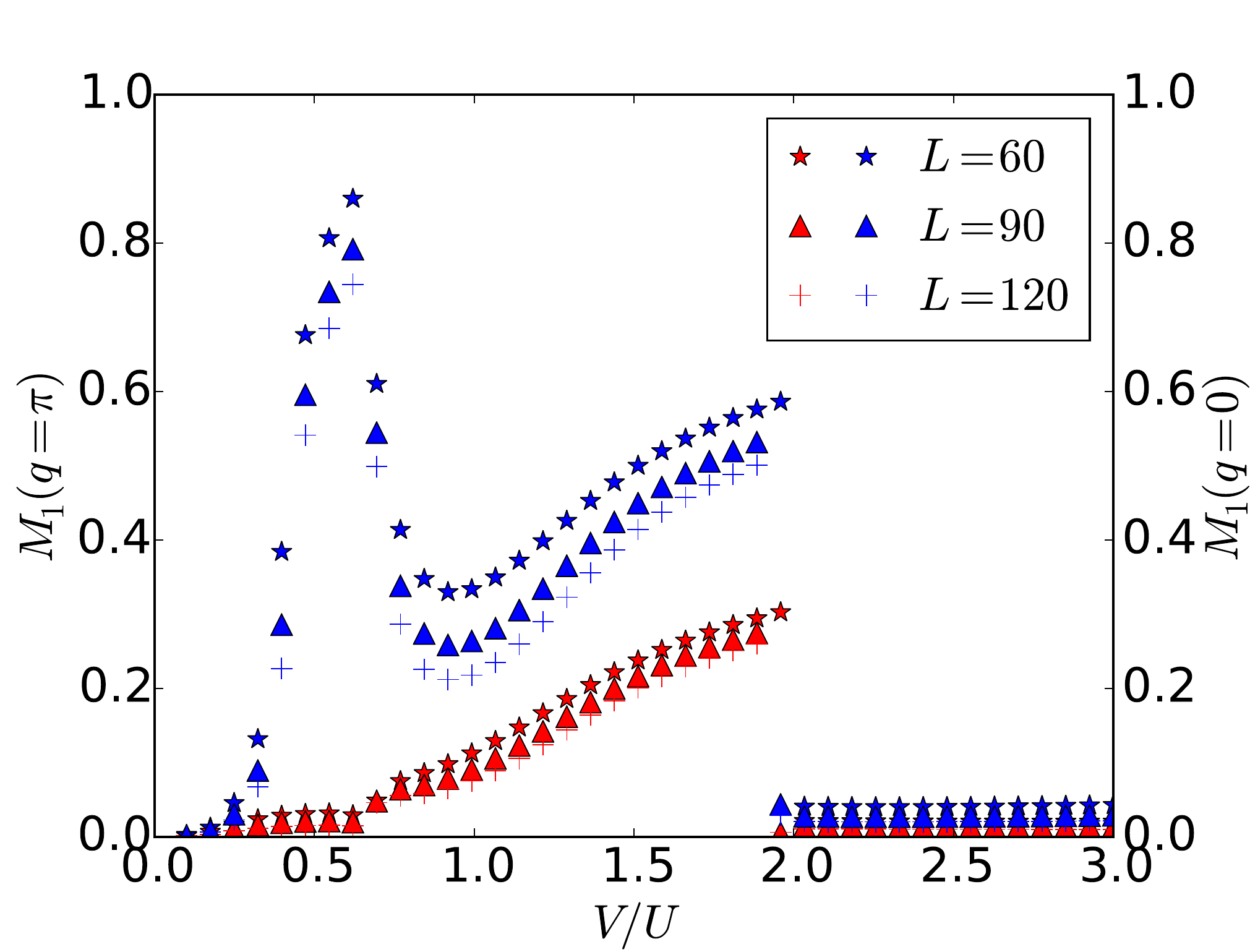}
	\end{minipage}\hfill\\
	\caption{(color online) Fourier components of the single particle off-diagonal correlations $M_1(q)$ (\ref{bcd}) at $q=0$ (red) and $q=\pi$ (blue) as a function of (a) $t/U$ for $V/U=0.5$ and of (b) $V/U$ for $t/U=0.02$. Different symbols correspond to different system sizes $L$ ($L=60,90,120$, see legenda), keeping $N=2L$. The discontinuity at $V/U\simeq 2$ is associated with the appearance of density modulations with quasi-momentum $q=2\pi/3$, see also Fig. \ref{fig:Struct:cut:nnn}.} 
	\label{fig:Mb:nnn}
\end{figure}

In the superfluid phase the spectrum of the single-particle off-diagonal correlations have non-vanishing Fourier component at $q=0$ and at $q=\pi$. Figure \ref{fig:Mb:nnn}(a) displays these Fourier components as a function of $t/U$ for $V=U/2$. The behavior is similar to the nearest-neighbor case, Fig. \ref{fig:5}(a). Now, however, the incompressible phase occurs on a substantially smaller interval of $t/U$ values. We attribute this effect to the next-nearest-neighbor terms of the interaction-induced tunneling. In fact, from Eq. (\ref{Tshiftnnn}) we can see that these terms tend to increase the effective hopping coefficient. 

\begin{figure}[h!]
	\centering
	\begin{minipage}[t]{0.45\textwidth}\vspace{0pt}
		\includegraphics[width=\textwidth]{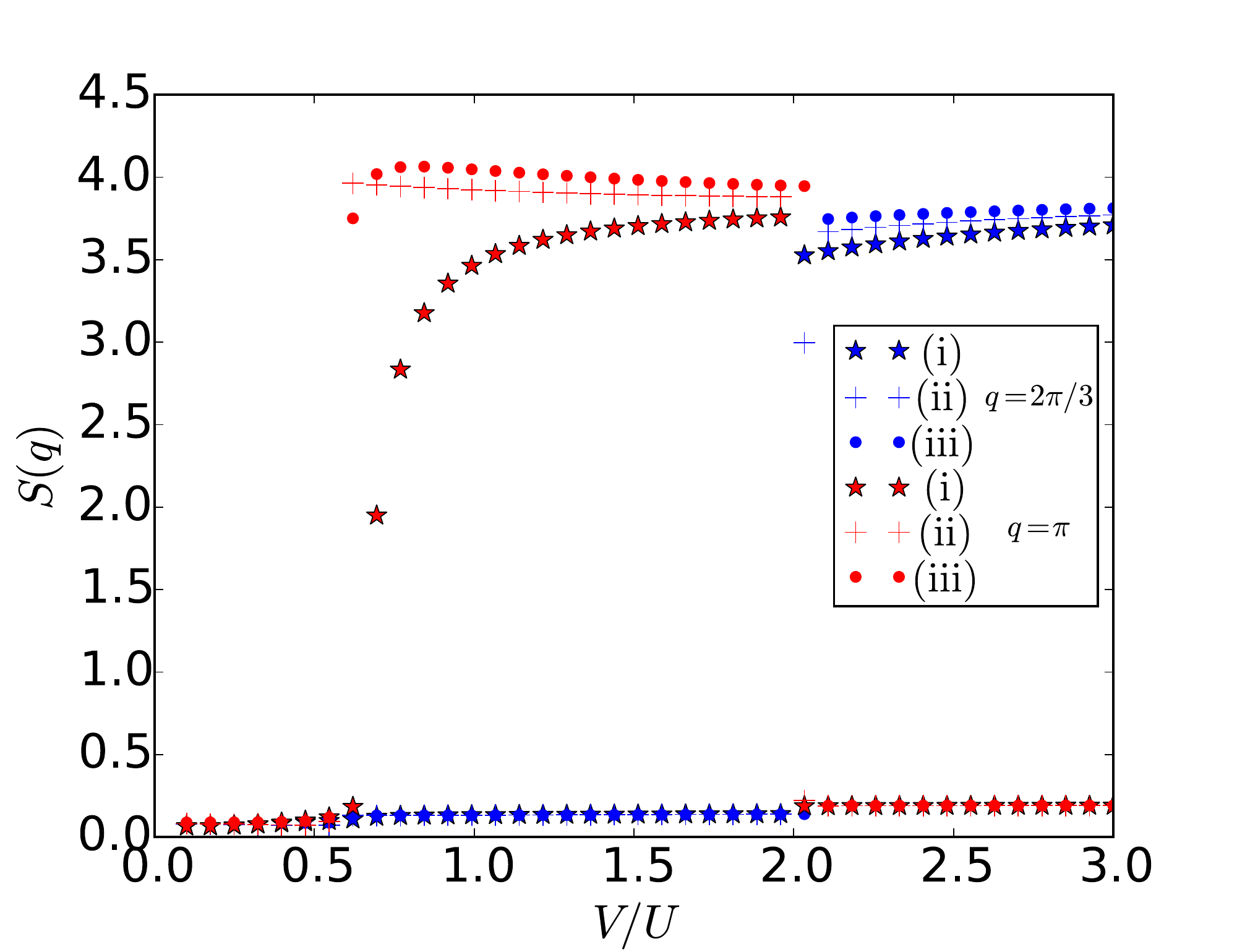}
	\end{minipage}\hfill\\
	\caption{(color online) The component at $q=\pi$ (red) and $q=2\pi/3$ (blue) of the structure form factor, Eq. \eqref{DW_orderparameter}, signalling the onset of density modulations. The data are taken at the sweeps (i), (ii), (iii) of Fig. \ref{fig:PD:nn}(d). Here, $t/U=0.02$ at (i), $t/U=0.157$ at (ii), and $t/U=0.25$ at (iii).} 
\label{fig:Struct:cut:nnn}
\end{figure}

\begin{figure}[h!]
	\centering	
	\begin{minipage}[t]{0.45\textwidth}\vspace{0pt}
		\includegraphics[width=\textwidth]{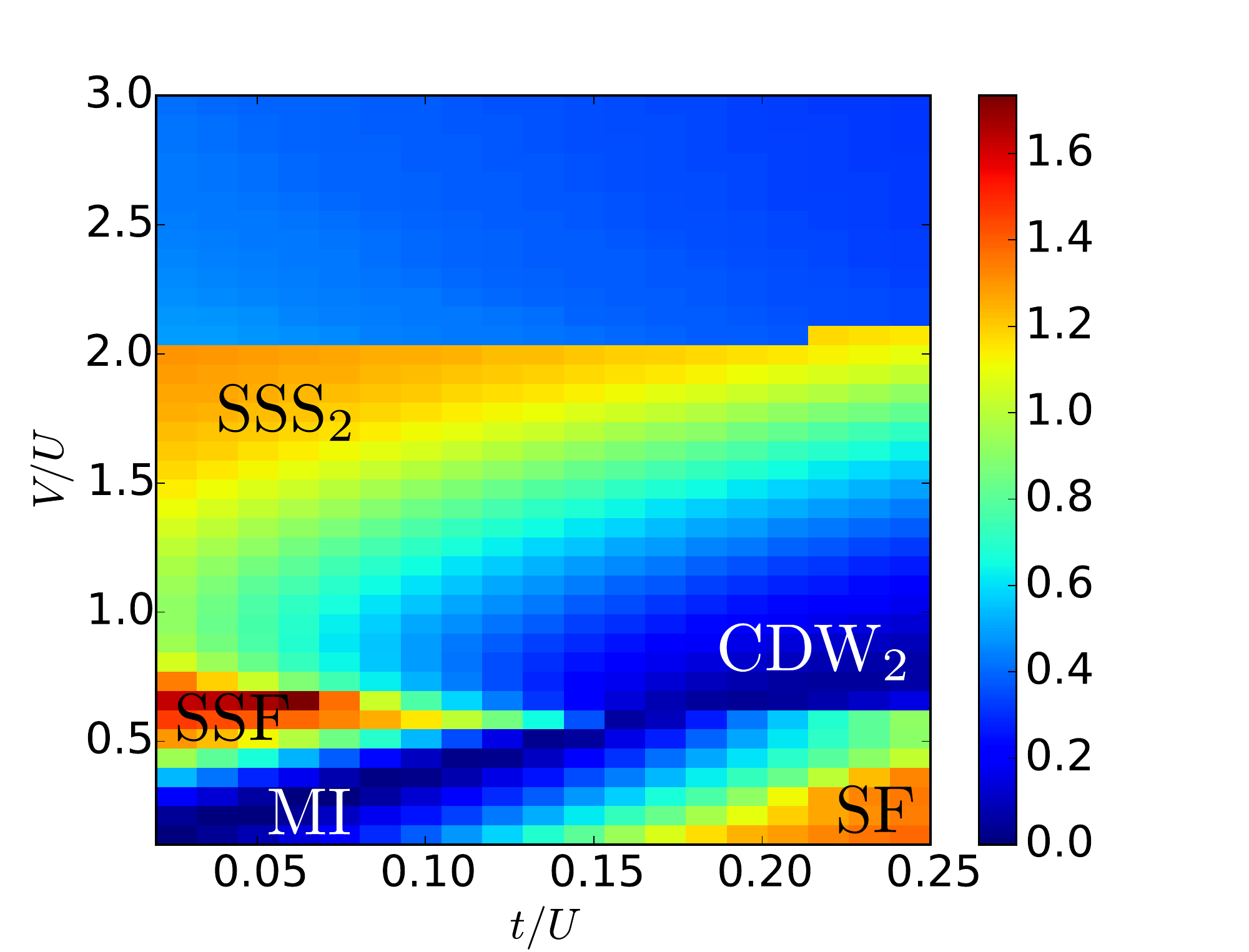}
	\end{minipage}\hfill\\
	\caption{(color online) Contour plot of the von Neumann entropy as a function of the ratio $V/U$ and of the ratio $t/U$ for the ground state of Eq. \eqref{HBH:nnn}. The phases are labeled according to the classification of Table \ref{Table:1}. The parameters are the same as in Fig. \ref{fig:PD:nn}, the size of the subsystem is $L_B=30$ sites.} 
	\label{fig:16b}
\end{figure}

The behavior of the Fourier components for $t\ll U$ is shown in Fig. \ref{fig:Mb:nnn}(b) as a function of $V/U$. For $V/U\lesssim 2$ it is similar to the nearest-neighbor model. Also in this case it exhibits the features of a continuous transition. The maximum of the $\pi$ component, however, is shifted to larger values (compare to Fig.~\ref{fig:6}), which is consistent with our preliminary considerations. Moreover, for $V/U$ to the right of the maximum, the slope with which both Fourier components at $q=0$ and $q=\pi$ increase is larger than for the nearest-neighbor interaction. 
At $V\sim 2U$ both components undergo an abrupt transition to a very small, non-vanishing value. At this point, the structure acquires a periodic density modulation at wave number $q=2\pi/3$, as visible from Fig. \ref{fig:Struct:cut:nnn}. The transition is thus discontinuous. The new phase seems to be a SSS$_3$. However, in the corresponding region the entanglement entropy, Fig. \ref{fig:16b}, takes very small values. Its nature shall be clarified by a future analysis for larger system sizes.

\section{Conclusions}
\label{Sec:5}

In this work we have characterised the Bose-Hubbard model of dipolar bosons in a one-dimensional lattice and in a parameter regime, where tunneling induced by interactions can interfere with the hopping due to the kinetic energy. We have found that significant effects of interaction-induced hopping are particularly important for sufficiently large densities. We then considered density $\rho=2$ in deep optical lattices and identified the parameter regime for which perfect destructive interference can occur. 

Quantum interference between correlated and single-particle tunneling qualitatively modifies the phase diagram. One important result is that it gives rise to an effective "interaction-induced atomic limit". This interaction-induced atomic limit is responsible for the appearance of an incompressible phase for finite values of the kinetic energy, where one otherwise expects superfluidity. Another important consequence of correlated tunneling is that at vanishing kinetic energy the dipolar interaction establishes superfluidity with a site-oscillating phase. This staggered superfluidity is separated from a staggered supersolidity by a continuous transition at $V\simeq U/2$. At this transition point the entanglement entropy exhibits a narrow peak.

We have compared the phase diagrams when the terms of the power-law interactions of the Bose-Hubbard model are truncated to (i) the first neareast neighbors and then (ii) to the next-neareast neighbors. Qualitative differences are visible at sufficiently large interaction strengths, where the next-nearest neighbor terms start to compete with the other terms. In particular, for next-nearest neighbor interactions, at larger dipolar strengths we have found discontinuous transitions to structures with larger Wigner-Seitz cells. 

The interference between single-particle and correlated hopping is a consequence of the behavior of the Bose-Hubbard coefficients as a function of the dipole moment. This interference cuts the phase diagram into two topologically different superfluid phases. In an experiment with given atomic species one would sweep along the line at fixed dipole moment (see Fig. \ref{app:fig:1} (a)). Then our results indicate that, by tuning the ratio $t/U$ one would observe either transitions from incompressible to SSF phases or to "normal" SF. Therefore, species with different dipole moments are  characterized by either SSF or "normal" SF phases. Moreover, our results indicate that special values of the dipole moment $d$ can exist, for which the gas remains always in the interaction induced atomic limit, independently of $t$. We note that these behaviors are also present at lower densities. Here, they become visible at larger values of the dipolar interactions, for which one shall consider the contribution of higher bands, as done for instance in Ref. \cite{Bissbort:2012, Lacki:2013, Major:2014}. Future studies will analyze the effect of correlated tunneling on the phases at incommensurate densities \cite{Chatterjee:2020} as well as at fractional densities,  where it might significantly affect the physics of Fibonacci anyon excitations  \cite{Duric:2016}.

\acknowledgements
We are especially grateful to George G. Batrouni and Kuldeep Suthar for helpful comments. We also thank Florian Cartarius, Anna Minguzzi, Simone Montangero, Luis Santos, Shraddha Sharma, and Ferdinand Tschirsich for discussions. This work has been supported by the German Research Foundation (the priority program No. 1929 GiRyd), by the European
Commission (ITN ColOpt) and by the German Ministry of Education and Research (BMBF) via the QuantERA projects NAQUAS and QTFLAG. Projects NAQUAS and QTFLAG have
received funding from the QuantERA ERA-NET Cofund in Quantum Technologies implemented within the European Union's Horizon 2020 program. The support by National Science Centre (Poland) under Project No. 2017/25/Z/ST2/03029 (J.Z.) is also acknowledged. 

\appendix

\section{Coefficients of the extended Bose-Hubbard Hamiltonian} \label{appendix:1}
In this appendix we introduce the integral expressions of the coefficient of the Bose-Hubbard Hamiltonian in Sec. \ref{Sec:2}. We also show how we determine the Hamiltonian parameters of Fig. \ref{appfig:1}. 

The extended Bose-Hubbard model we consider is obtained by inserting Eq. (\ref{exp}) into the Hamiltonian of Eq. (\ref{secquant}):
\begin{align}
\hat{H}=-\sum_{i,j}t_{i,j} \hat{a}^\dagger_i\hat{a}_j +\sum_{i,j,k,l} V_{i,j,k,l} \hat{a}^\dagger_i \hat{a}^\dagger_j \hat{a}_k \hat{a}_l \ .
\end{align}
The tunneling coefficients are given by the integrals
\begin{align}    
\label{app:t}                 
t_{i,j} = \int_{-L/2a}^{L/a}dx \ w_i(x)\left( \frac{\hbar^2}{2m} \frac{\partial^2}{\partial x^2}- V_0\sin(\pi x /a)\right) w_{j}(x) \, ,
\end{align}
\noindent where $w_j(x)$ is the real-valued Wannier function. We define $t=t_{i,i+1}$ and $t_{\text{NNN}}=t_{i,i+2}$ and discard higher order terms.  
The interaction coefficients are defined by the expressions
\begin{align}
	\label{Vijkl}
	V_{i,j,k,l} =& \frac{1}{2}\int \int d\mathbf{r}_1  d\mathbf{r}_2  w_i(x_1)w_j(x_2)\times \nonumber \\
	\times &U_{\text{int}}(\mathbf{r}_1-\mathbf{r}_2)w_k(x_2)w_{l}(x_1)\Phi_0(y_1,z_1;y_2,z_2)\, ,
\end{align}
\noindent where $U_{\rm int}(\mathbf{r})=U_g(\mathbf{r})+U_\alpha(\mathbf{r})$ and $\Phi_0\equiv|\phi_0(y_1,z_1)|^2|\phi_0(y_2,z_2)|^2$, see Sec. \ref{Sec:2}. 
The coefficients we use in Sec. \ref{Sec:2} are connected to the integral expression in Eq. (\ref{Vijkl}) as follows: \\
The onsite interaction present in Eq. (\ref{H:BH}) is given by $U=2V_{i,i,i,i}$. 
The coefficients of the extended nearest neighbor Bose-Hubbard Hamiltonian, Eq. (\ref{HBH:nn}), have the form 
\begin{eqnarray}
&&V = 2\left( V_{i,i+1,i+1,i}+V_{i,i+1,i,i+1} \right)\,,\\
&&T= -\left(V_{i,i,i+1,i}+V_{i,i,i,i+1}\right)\,,\\
&&P= 2V_{i,i,i+1,i+1}\, .
\end{eqnarray}
These coefficients include the overlap integrals of Wannier functions of nearest neighboring sites. 
The expression of the next-nearest neighbor interaction coefficients in Eq. (\ref{HBH:nnn}) are:
\begin{eqnarray}
&& V_{\text{NNN}}=2\left( V_{i,i+2,i,i+2}+V_{i,i+2,i+2,i} \right)\,,\\
&& T_{\text{NNN}}^1= -2\cdot\left(V_{i+2,i,i+1,i}+V_{i+2,i,i,i+1}\right)\,,\\
&& T_{\text{NNN}}^2=-2\cdot\left(V_{i+2,i,i+2,i+1}+V_{i+2,i+1,i,i+2}\right)\,,\\
&&T_{\text{NNN}}^3=-2\cdot \left(V_{i+2,i+1,i+1,i}+V_{i+2,i+1,i,i+1}\right)\,,\\
&&T_{\text{NNN}}=-\left(V_{i,i,i+2,i}+V_{i,i,i,i+2}\right)\,,\\
&&P_{\text{NNN}}^1=2\left(V_{j+2,j+1,j,j}+V_{j+1,j+2,j,j} \right)\,,\\
&&P_{\text{NNN}}^2=2\left(V_{j+2,j,j+1,j+1}+V_{j,j+2,j+1,j+1}\right)\,,\\
&&P_{\text{NNN}}^3=2\left(V_{j+2,j+2,j+1,j}+V_{j+2,j+2,j,j+1} \right)\,.
\end{eqnarray}
We determine the Hamiltonian parameters as follows. We first decompose the coefficient $V_{i,j,k,l}$ as 
\begin{align*}
V_{i,j,k,l} = V^\alpha_{i,j,k,l}+V^g_{i,j,k,l} \ ,
\end{align*}
where the contribution of the contact interaction to the overall coefficient is given by
\begin{align}
\label{Vijklg}
V^g_{i,j,k,l} = \frac{g}{2} \int dx w_{i}(x)w_{l}(x)w_j(x)w_k(x) \ .
\end{align}
The coefficients due to the power law interactions are then calculated by means of the convolution method \cite{Wall2013}:
\begin{align}
\label{conVijkl}
&V^\alpha_{i,j,k,l}=\frac{1}{2}\int dx dy w_{i}(x)w_{l}(x)\Phi(y) \nonumber\\
&\times \mathcal{F}_\mathbf{k}^{-1}\left[ \tilde{V}^\alpha_{2D}(\mathbf{k})\mathcal{F}_{\mathbf{k}}\left[w_{j}(x')w_{k}(x')\Phi(y')\right]\right] \ ,
\end{align}
where $\mathcal{F}_{\mathbf{k}}$ is the Fourier transform from position to momentum space. Here $\Phi(y)=\frac{1}{\sqrt{\pi}\sigma}e^{-y^2/\sigma^2}$ is the probability density of the ground state of the harmonic trap along the $y$-direction, the width is $\sigma=\sqrt{\hbar/m \omega}$ and it is the same for the $y$- and the $z$-direction.
In Eq. (\ref{conVijkl}) the expression $\tilde{V}^\alpha_{2D}(\mathbf{k})$ is the effective interaction in momentum space and reads as\cite{cartarius:2017}
\begin{align}
\tilde{V}_{2D}(k_y,k_x) = \frac{C_{dd}}{2\sigma} \left[\frac{2}{3}\sqrt{\frac{2}{\pi}}-q\sigma \ \text{erfc}\left(\frac{\sigma q}{\sqrt{2}}\right)\right] \ , \label{2deff}
\end{align}
where $q^2=k_x^2+k_y^2$. 
The expression in Eq. \eqref{2deff} is the effective 2D interaction in momentum space, where we integrate out the $z$-coordinate. For further details see Ref. \cite{cartarius:2017}. We calculate numerically the integral in Eq. (\ref{Vijklg}) and (\ref{conVijkl}) as a function of $d$ and $a_s$. In our calculations the lattice and trap parameters are kept constant and take the values $V_0=8E_R$ and $\sigma/a=1/\pi\sqrt[4]{50}$.

\section{Details on the numerical implementation}
\label{App:DMRG}
Our results are obtained with a DMRG numerical program, where we make use of the ITensor C++ library for implementing tensor network calculations \cite{itensor}. In our simulations we use a maximum bond dimension of $\beta=600$. The cutoff $\epsilon$ is set to $\epsilon = 10^{-12}$, which determines the number of singular values discarded after each singular value decomposition (SVD) step. The energy error goal is set to $\epsilon_{\text{goal}}=10^{-16}$ and the maximum number of particles per site is fixed to $n_{\text{max}} = 10$. 
We also add a boundary term $\hat{H}_{ad} = \floor{2\rho}\left(Vn_1+V_{\text{NNN}}n_2\right)$ in order to lift the degeneracy in the CDW$_j$ phases and the Haldane phase. 
To improve the convergence we run the simulation for four different initial states: the CDW$_j$ states $\ket{\Phi}_{\text{init}}=\otimes_{k} \ket{j\cdot\rho}_k \otimes_{l} \ket{0}_l$ with $k \in \left\lbrace \mathbb{A}={j\cdot m|m\in \mathbb{N}} \right\rbrace$ and $l \in \mathbb{N}\backslash \mathbb{A}$, the MI state $\ket{\Phi}_{\text{init}}=\otimes_{k=1}^L\ket{\rho}_k$ and a random initial state. The random state is a superposition of Fock states $\ket{\Phi}_{\text{init}}=\frac{1}{\sqrt{n_{\text{iter}}}}\sum^{n_{\text{iter}}}_k \left(\otimes_i\ket{n_i}\right)_k$, where $n_i \in \mathbb{N}$ is chosen randomly out of the interval $[0,n_{\text{max}}]$ with the constrain $\sum_{i=1}^Ln_i=\rho$. We choose the number of superimposed Fock state to be $n_{\text{iter}}=100$. 
At the end of the simulation we identify the ground state with the state at lowest energy.	
In order to eliminate the boundary effects we determine the expectation values over the ground state by reducing the chain length by $n_{\text{sit}}$ on each edge in order to eliminate boundary effects. We choose $n_{\text{sit}}=10$. 

In order to calculate the von Neuman entropy introduced in Sec. \ref{Sec:4} we split the system into subsystem A and B. We then perform at the bond of these two subsystems A and B a singular value decomposition (SVD) of the final ground state coefficients. We determine the von Neuman entropy using the singular values $s_\alpha$ given by the eigenvalues of the diagonal $S$-matrix of the SVD \cite{itensor,Schollwoeck2011}
\begin{equation}
S_{\text{vN}}= -\sum_{\alpha}s_{\alpha}^2\ln \left( s_{\alpha}^2\right) \label{entangle} \ .
\end{equation} Here we choose the length of the subsystem A to be half of the length of the system.

\section{Supplementary details on the ground state phase diagram for nearest-neighbor couplings}
\label{App:PD:Details}
\begin{figure}[h!]
	\centering
	\begin{minipage}[t]{0.4\textwidth}\vspace{0pt}
		\includegraphics[width=\textwidth]{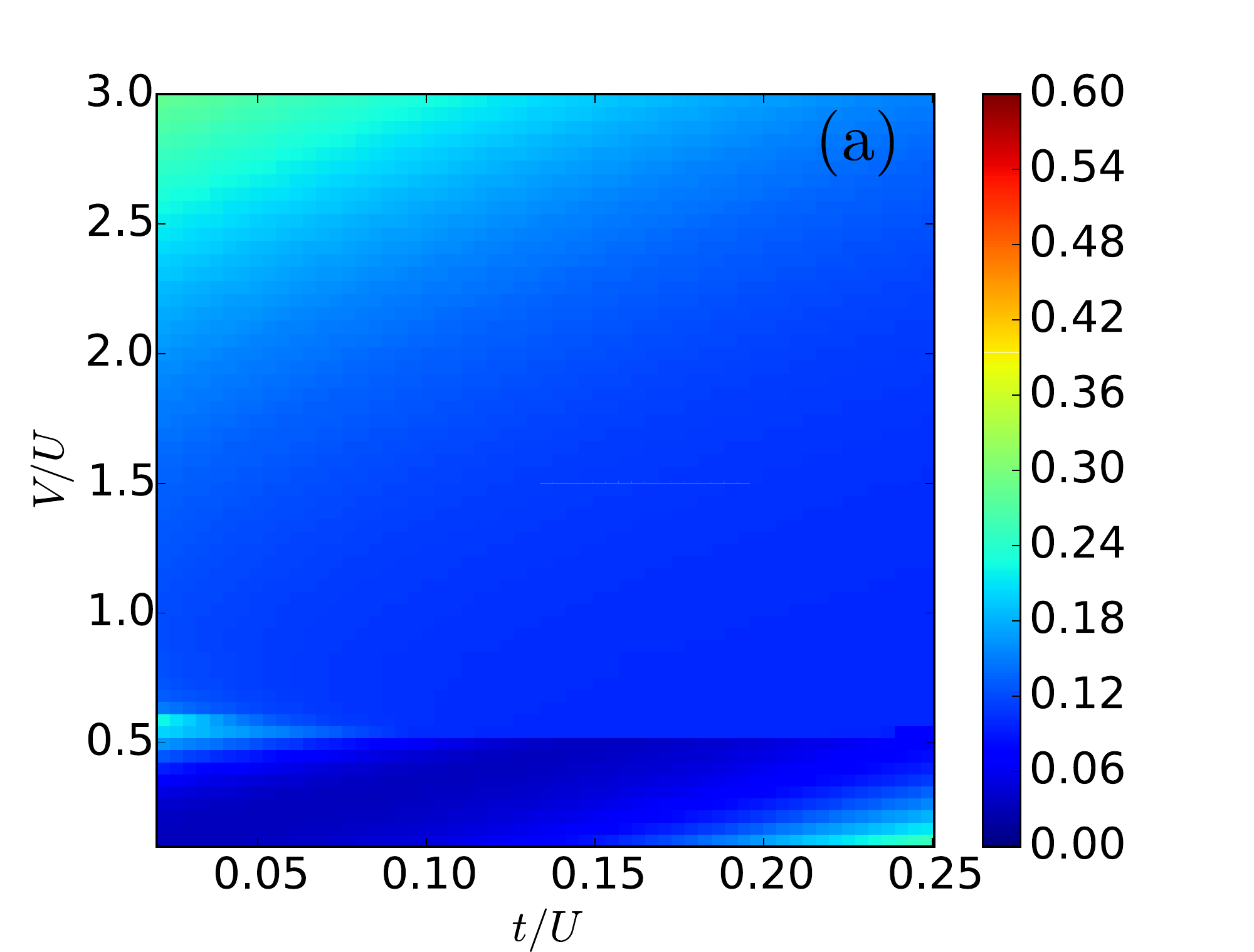}
	\end{minipage}\hfill\\
	\begin{minipage}[t]{0.4\textwidth}\vspace{0pt}
		\includegraphics[width=\textwidth]{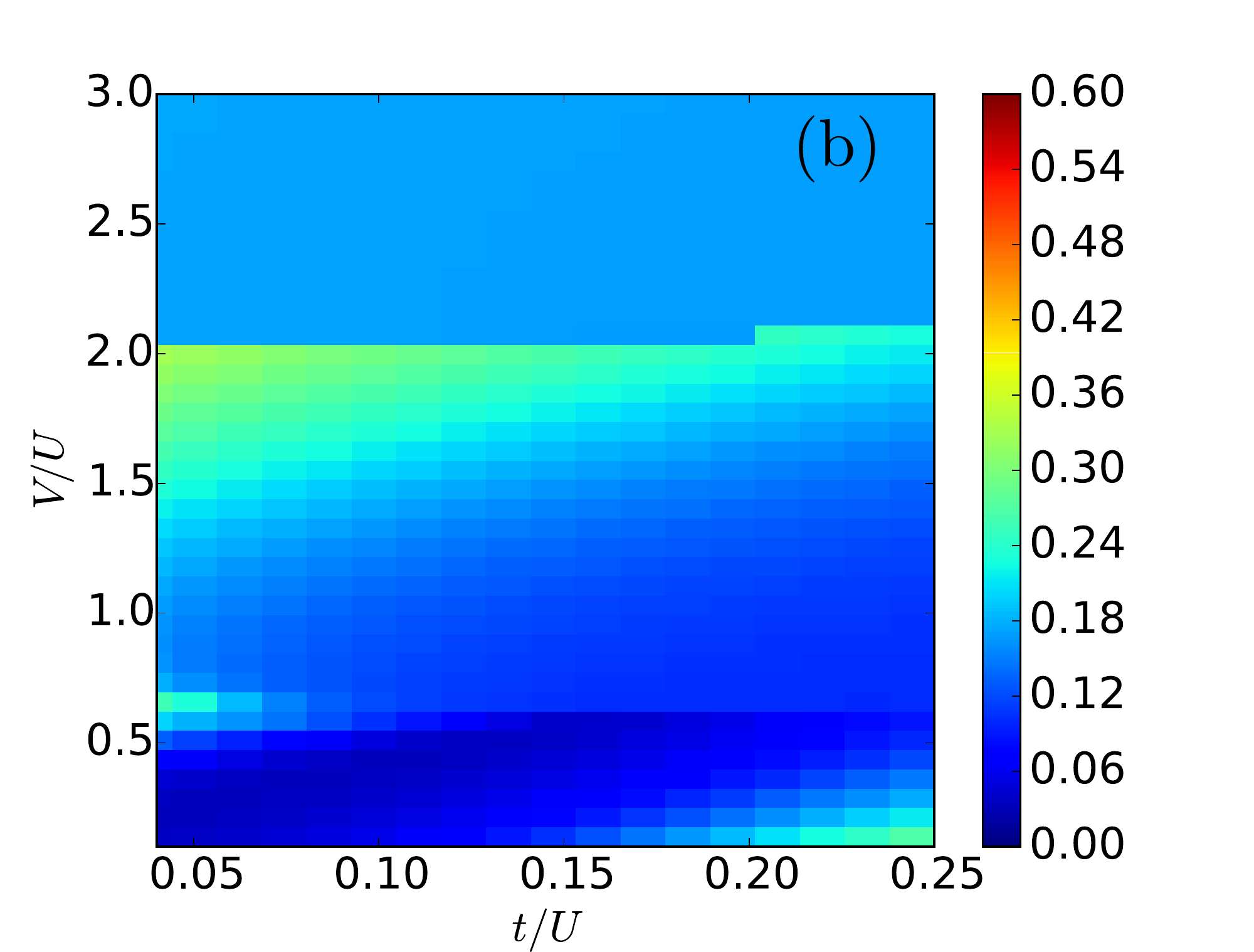}
	\end{minipage}\hfill\\
	\caption{(color online) Contour plot of the pair condensate density $M_2(q=0)$ (\ref{pcd}) as a function of the nearest neighbor interaction strength $V$ and the tunneling rate $t$ both in units of the on-site interaction strength $U$ for the (a) nearest-neighbor and (b) next-nearest-neighbor model. The number of particles is fixed to $N=120$ and the number of lattice sites is given by $L=60$. } 
	\label{app:fig:1}
\end{figure}

Here we provide additional details on the results of Sec. \ref{Sec:4}.
We checked the presence of a pair superfluid phase over the whole parameter range by looking at the Fourier transform of the pair correlations $M_2(q)$ given by Eq. (\ref{pcd}). We found nonvanishing Fourier components of $M_2(q)$ only at $q=0,\pi$. Moreover, when $M_2(q)$ has nonvanishing components, then we always find that $M_2(q=0)> M_2(q=\pi)$. Figure  \ref{app:fig:1} displays the contour plot of the Fourier transform at $q=0$ in the $V/U-t/U$ parameter plane for (a) the nearest-neighbor and (b) the next-nearest neighbor case. We note that the contour plot of the Fourier transform at $q=\pi$ is finite if $M_2(q=0)$ and $S(q=\pi)$ are both finite.

The behavior of observable $M_2(q=0)$ across the phase diagram follows the behavior of the Fourier transform of the single particle correlations, see Fig. \ref{fig:PD:nn} and \ref{fig:PD:nnn}. Nevertheless, where they are finite, the Fourier components $M_2(q)$ are always smaller than the corresponding Fourier components of the single particle correlations. We conclude that there is no PSF in the parameter regime we considered. 

\begin{figure}[h!]
	\centering
	(a)
	\begin{minipage}[t]{0.4\textwidth}\vspace{0pt}
		\includegraphics[width=\textwidth]{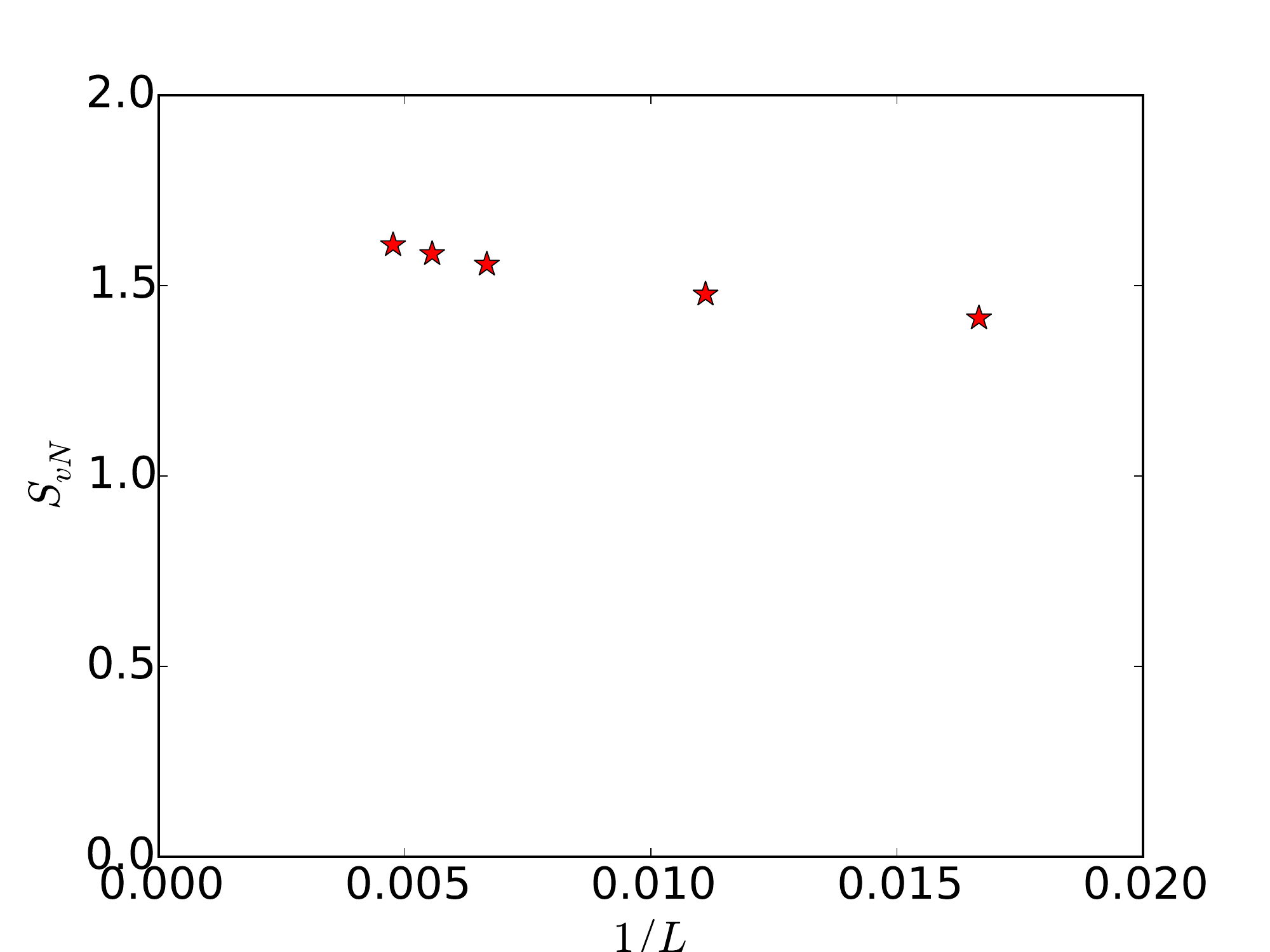}
	\end{minipage}\hfill\\
	(b)
	\begin{minipage}[t]{0.4\textwidth}\vspace{0pt}
		\includegraphics[width=\textwidth]{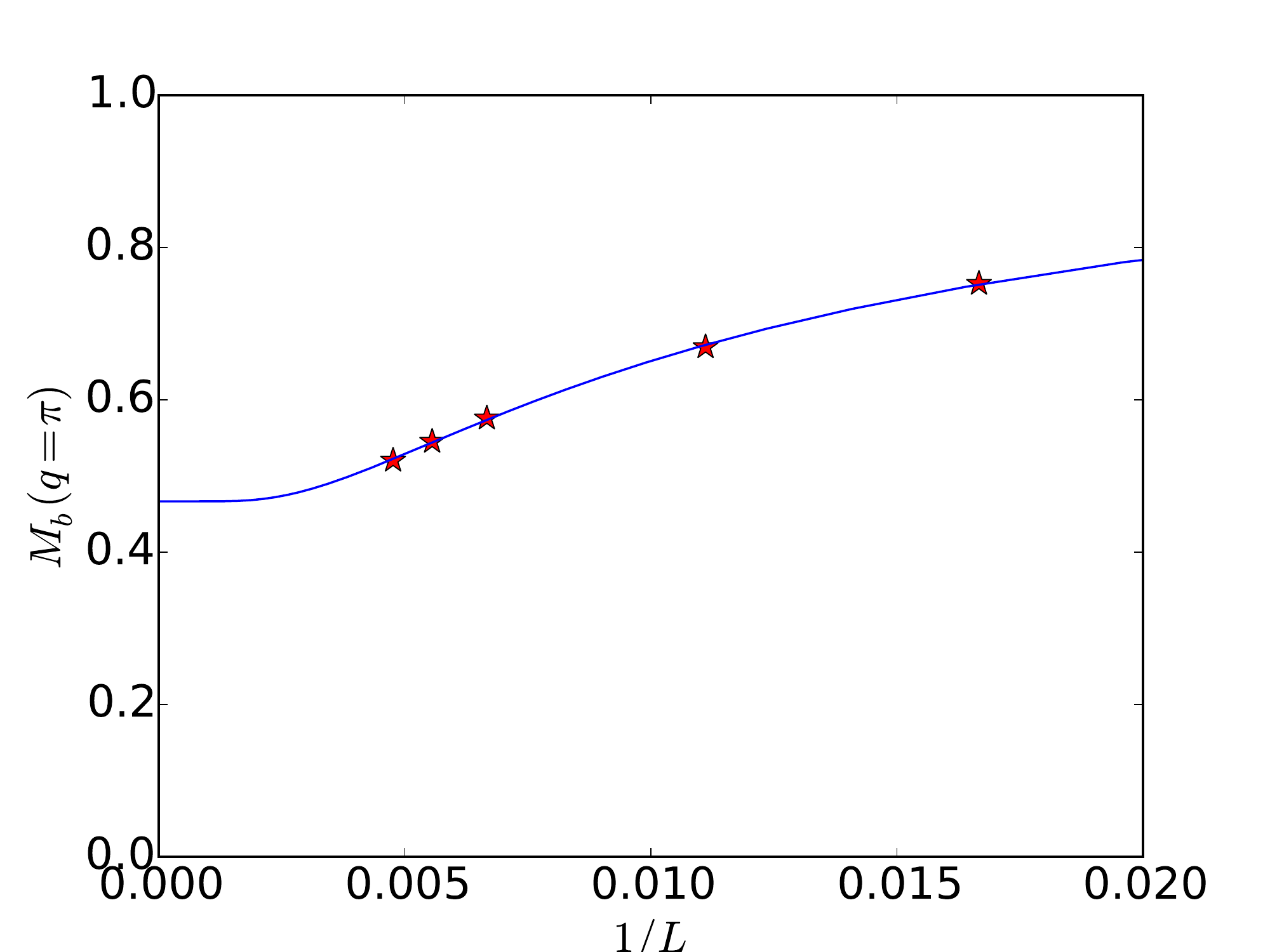}
	\end{minipage}\hfill\\
	\caption{(color online) (a) The von Neuman entropy (\ref{entangle}) and (b) the Fourier transform of the single particle correlations $M_1(q)$ (\ref{bcd}) at $q=\pi$ at a tunneling rate of $t/U=0.04$ and a nearest neighbor interaction strength of $V/U=0.5$ as a function of one over the number of lattice sites $L$. The blue curve in (b) shows the exponential function $F(L)=0.45\exp(-0.01(L+18.64))+0.47$.  } 
	\label{app:fig:2}
\end{figure}

In Sec. \ref{Sec:4} we observe the presence of a staggered superfluid phase around $V/U\sim 0.5$ and for small $t/U$. Here we check the presence of this phase in the thermodynamic limit. Therefore Fig. \ref{app:fig:2} shows in subplot (a) the von Neuman entropy (\ref{entangle}) and in subplot (b) the Fourier transform of the single particle correlations $M_1(q)$, see Eq. (\ref{bcd}), at $q=\pi$ as a function of one over the number of lattice sites $1/L$ for a fixed $t/U=0.04$ and $V/U=0.5$. 
By inspecting Fig. \ref{app:fig:2} we can identify a convergence of the observables with increasing system size. By fitting the curve in Fig. \ref{app:fig:2} (b) with an exponential function and taking the limit for $L$ going to infinity we get a limit of $M_1(\pi)(L\rightarrow \infty)=0.47$ . We conclude that the staggered superfluid phase around $V/U\approx 0.5$ and for small $t/U$ is present in the thermodynamic limit. 

\begin{figure}[h!]
	\centering
	\begin{minipage}[t]{0.45\textwidth}\vspace{0pt}
		\includegraphics[width=\textwidth]{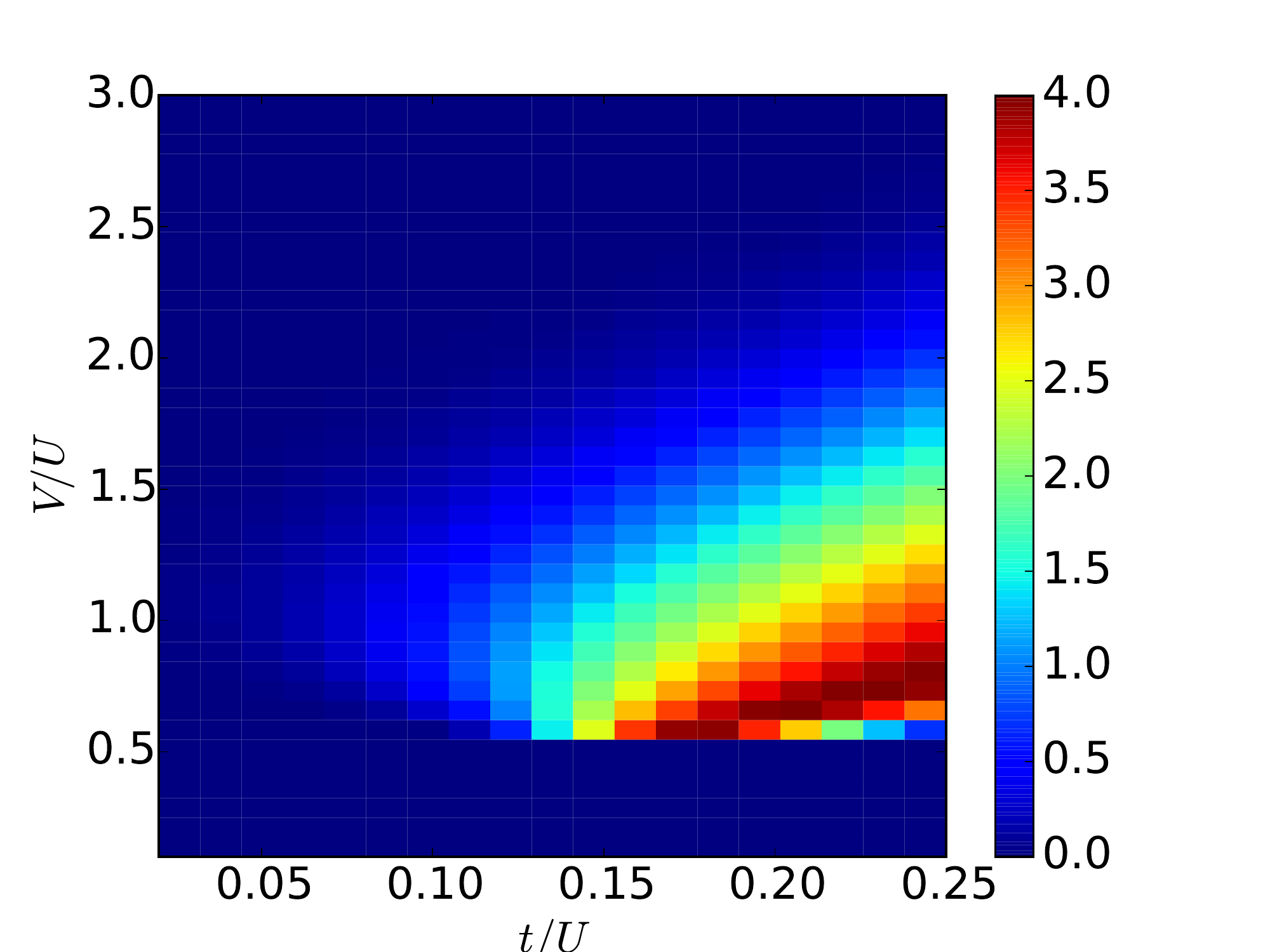}
	\end{minipage}\hfill\\
	\caption{(color online) Contour plot of the string order parameter, Eq. \eqref{eq:string}, for the ground state of Eq. \eqref{HBH:nn}, as a function of the ratio $V/U$ and $t/U$. The parameters are the same as in Fig. \ref{fig:PD:nn}.} 
	\label{app:fig:3}
\end{figure}

Figure \ref{app:fig:3} displays the contour plot of the string order parameter given by Eq. (\ref{eq:string}. For the calculation of the string order parameter we choose $r=L/2$ and discard the outer $L/4$ sites on both sides of the chain \cite{Rossini2012}. 

By comparing the string order parameter and the structure form factor in Fig. \ref{fig:PD:nn}(d) we cannot identify any region, where the string order parameter is finite and the structure form factor vanishes.


\begin{thebibliography}{10}
\bibitem{RMP:Bloch-Dalibard-Zwerger}
I. Bloch, J. Dalibard, and W. Zwerger,
Rev. Mod. Phys. \textbf{80}, 885 (2008).


\bibitem{Fisher1989}
M. P. A. Fisher, P. B. Weichman, G. Grinstein, and D. S. Fisher,
Phys. Rev. B \textbf{40}, 546 (1989). 


\bibitem{Greiner2002}
M. Greiner, O. Mandel, T. Esslinger, T. W. H\"ansch, and I. Bloch,
Nature \textbf{415}, 39 (2002).




\bibitem{Boettcher2019}
F. B\"ottcher, J.-N. Schmidt, M. Wenzel, J. Hertkorn, M. Guo, T. Langen, and T. Pfau,
Phys.~Rev.~X \textbf{9}, 011051 (2019).

\bibitem{Chomaz2019}
L. Chomaz, D. Petter, P. Ilzh\"ofer, G. Natale, A. Trautmann, C. Politi, G. Durastante, R. M. W. van Bijnen, A. Patscheider, M. Sohmen, M.J. Mark, and F. Ferlaino,
Phys.~Rev.~X \textbf{9}, 021012 (2019).



\bibitem{Tanzi:2019}
L.~Tanzi, E.~Lucioni, F.~Fam\'a, J.~Catani, A.~Fioretti, C.~Gabbanini, R.~N.~Bisset, L.~Santos, and G.~Modugno,
Phys. Rev. Lett. \textbf{122}, 130405 (2019).














%
%
%
%



\bibitem{Je}
S. A. Moses, J. P. Covey, M. T. Miecnikowski, D. S. Jin, and
J. Ye, Nat. Phys. {\bf 13}, 13 (2017).



\bibitem{dePaz:2013}
A. de Paz, A. Sharma, A. Chotia, E. Maréchal, J. H. Huckans, P. Pedri, L. Santos, O. Gorceix, L. Vernac, and B. Laburthe-Tolra,
Phys. Rev. Lett. \textbf{111}, 185305 (2013).


\bibitem{Moses:2015}
S. A. Moses, J. P. Covey, M. T. Miecnikowski, B. Yan, B. Gadway, J. Ye, and D. S. Jin, 
Science, \textbf{350}, 659 (2015).


\bibitem{Reichsoellner:2017}
L. Reichs\"ollner, A. Schindewolf, T. Takekoshi, R. Grimm, and H.-Ch. Nägerl,
Phys. Rev. Lett. \textbf{118}, 073201 (2017).








\bibitem{Covey:2016}
J. P. Covey, S. A. Moses, M. G\"arttner, A. Safavi-Naini, M. T. Miecnikowski, Z. Fu, J. Schachenmayer, P. S. Julienne, A. M. Rey, D. S. Jin, and J. Ye,
Nat. Commun. \textbf{7}, 11279 (2016).



\bibitem{Baier2016}
S.~Baier, M.~J.~Mark, D.~Petter, K.~Aikawa, L.~Chomaz, Z.~Cai, M.~Baranov, P.~Zoller, and F.~Ferlaino,
Science\ \textbf{352}, 201 (2016).



%
%
%
%



\bibitem{Lahaye}
T. Lahaye, C. Menotti, L. Santos, M. Lewenstein, and T. Pfau,
Rep. Prog. Phys. {\bf 72}, 126401 (2009).

\bibitem{Pollet:2010}
L. Pollet, J. D. Picon, H. P. B\"uchler, and M. Troyer, 
Phys. Rev. Lett. \textbf{104}, 125302 (2010).



\bibitem{Pupillo}
M. A. Baranov, M. Dalmonte, G. Pupillo, and P. Zoller, Chemical Reviews {\bf 112}, 5012 (2012).


\bibitem{Pupillo:2010}
B. Capogrosso-Sansone, C. Trefzger, M. Lewenstein, P. Zoller, and G. Pupillo,
Phys. Rev. Lett. \textbf{104}, 125301 (2010).


\bibitem{Menotti2007}
C. Menotti, C. Trefzger, and M. Lewenstein,
Phys. Rev. Lett. \textbf{98}, 235301 (2007).


\bibitem{Goral2002}
K. Goral, L. Santos, and M. Lewenstein
Phys. Rev. Lett. \textbf{88}, 170406 (2002).



\bibitem{Dutta2015}
O.~Dutta, M.~Gajda, P.~Hauke, M.~Lewenstein, D.-S.~L\"uhmann, B.~A.~Malomed, T.~Sowi\'nski, and J. Zakrzewski,
Rep. Prog. Phys. \textbf{78}, 066001 (2015).

\bibitem{Yi:2007}
S. Yi, T. Li, and C. P. Sun,
Phys. Rev. Lett. \textbf{98}, 260405 (2007).


\bibitem{Sinha:2005}
D. L. Kovrizhin, G.V. Pai, and S. Sinha,
Europhys. Lett. \textbf{72}, 162 (2005). 



%
%
%
%









\bibitem{Kuehner:1999}
T.~D.~K\"uhner, S.~R.~White, and H.~Monien,
Phys.~Rev.~B \ \textbf{61}, 12474 (2000).


\bibitem{Sengupta2005}
P.~Sengupta, L.~P.~Pryadko, F.~Alet, M.~Troyer, and G.~Schmid,
Phys.~Rev.~Lett.\ \textbf{94}, 207202 (2005).


\bibitem{Batrouni2006}
G.~G.~Batrouni, F.~H\'ebert, and R.~T.~Scalettar,
Phys.~Rev.~Lett.\ \textbf{97}, 087209 (2006).

\bibitem{Mishra:2009}  
T. Mishra, R. V. Pai, S. Ramanan, M. S. Luthra, and B. P. Das,
Phys. Rev. A \textbf{80}, 043614 (2009).

\bibitem{Otterlo:1994}
A. van Otterlo, and K. H. Wagenblast,
Phys. Rev. Lett. \textbf{72}, 3598 (1994).

\bibitem{Batrouni1995}
G.G. Batrouni, R.T.Scalettar, G.T.Zimanyi, and A. P. Kampf,
Phys. Rev. Lett. \textbf{74}, 2527 (1995).





\bibitem{DallaTorre2006}
E.~G.~Dalla Torre, E.~Berg, and E.~Altman,
Phys. Rev. Lett. \textbf{97}, 260401 (2006).

\bibitem{Deng2011}
X.~Deng, and L.~Santos,
Phys.~Rev.~B \textbf{84}, 085138 (2011).

\bibitem{Kawaki:2017}
K. Kawaki, Y. Kuno, and I. Ichinose,
Phys. Rev. B \textbf{95}, 195101 (2017).

\bibitem{Batrouni2013}
G.G. Batrouni, R.T. Scalettar, V. G. Rousseau, and B. Gr\'emaud,
Phys. Rev. Lett. \textbf{110}, 265303 (2013).

\bibitem{Rossini2012}
D.~Rossini, and R.~Fazio,
New~J.~Phys.~\textbf{14}, 065012 (2012).










\bibitem{Santos:2000}
L.~Santos, G.~V.~Shlyapnikov, P.~Zoller, and M.~Lewenstein,
Phys. Rev. Lett. \textbf{85}, 1791 (2000).

\bibitem{cartarius:2017}
F.~Cartarius, A.~Minguzzi, and G.~Morigi,
Phys.~Rev.~A\ \textbf{95}, 063603
(2017).

\bibitem{Goral:2002}
K.~Goral, and L.~Santos,
Phys. Rev. A \textbf{66}, 023613 (2002).

\bibitem{Schmidt}
K.P. Schmidt, J. Dorier, A. L\"auchli and F. Mila,
Phys. Rev. B {\bf 74}, 174508 (2006); K.P. Schmidt, J. Dorier, A. L\"auchli and F. Mila,
Phys. Rev. Lett. {\bf 100}, 090401 (2008).

\bibitem{sowinski2012dipol}
T.~Sowi\'nski, O.~Dutta, P.~Hauke, L.~Tagliacozzo and M.~Lewenstein,
Phys.~Rev.~Lett.\ \textbf{108}, 115301
(2012).

\bibitem{biedron2018extended}
K.~Biedro\'n, M.~\L\k{a}cki, and J.~Zakrzewski,
Phys.~Rev.~B\ \textbf{97}, 245102
(2018).

\bibitem{Amico:2010}
L. Amico, G. Mazzarella, S. Pasini, and F.S. Cataliotti,
New J. Phys. \ \textbf{12}, 013002 (2010).

\bibitem{Schollwoeck2011}
U. Schollw\"ock, 
Ann.~Phys. \textbf{326}, 96 (2011).

\bibitem{itensor}
\mbox{ITensor Library}, http://itensor.org.




\bibitem{Johnstone19}
D. Johnstone, N. Westerberg, C. W. Duncan and, P. \"Ohberg,
Phys.~Rev.~A\ \textbf{100}, 043614 (2019).


\bibitem{Astrakharchik2007}
G. E. Astrakharchik, J. Boronat, I. L. Kurbakov, and Yu. E. Lozovik,
Phys.~Rev.~Lett. \textbf{98}, 060405 (2007).





\bibitem{Jaksch:1999}
D. Jaksch, C. Bruder, J. I. Cirac, C. W. Gardiner, and P. Zoller,
Phys. Rev. Lett. {\bf 81}, 3108 (1998).


\bibitem{jakub2013}
M.~Maik, P.~Hauke, O.~Dutta, M.~Lewenstein, and J.~Zakrzewski, 
New J. Phys. \ \textbf{15}, 113041 (2013).







\bibitem{Juergensen2014}
O.~J\"urgensen, F.~Meinert, M.~J.~Mark, H.-Ch. N\"agerl, and D.-S. L\"uhmann,
Phys.~Rev.~Lett.~ \textbf{113}, 193003 (2014).


\bibitem{Luehmann2012}
D.-S.~L\"uhmann, O.~J\"urgensen, and K.~Sengstock, 
New~J.~Phys.~\textbf{14}, 033021 (2012).







\bibitem{Juergensen2015}
O. J\"urgensen, K. Sengstock, and D. S. L\"uhmann, Sci. Rep. \ \textbf{5}, 12912 (2015).










\bibitem{Local:Compressibility}
S. Wessel, F. Alet, M. Troyer, and G.G. Batrouni,
Phys. Rev. A \textbf{70}, 053615 (2004).

\bibitem{Roscilde2009}
T. Roscilde, New J. Phys. {\bf 11}, 023019 (2009).

\bibitem{Delande2009}
D. Delande, and J. Zakrzewski, Phys. Rev. Lett. {\bf 102}, 085301 (2009).

\bibitem{Jiang2012pair}
H.C.~Jiang, L.~Fu, and C.~Xu,
Phys.~Rev.~B\ \textbf{86}, 045129
(2012).

\bibitem{Luehmann2016}
D.-S. L\"uhmann, Phys.~Rev.~A \ \textbf{94}, 011603(R) (2016).

\bibitem{Dutta2011}
O.~Dutta, A.~Eckardt, P.~Hauke, B.~Malomed, and M.~Lewenstein, 
New J. Phys. \ \textbf{13}, 023019 (2011).

\bibitem{Qin:2003}
Sh.~Qin, J.~Lou, L.~Sun, and Ch.~Chen, 
Phys. Rev. Lett. \textbf{90}, 067202 (2003).

\bibitem{Knap:2012}
S. Ejima, H. Fehske, F. Gebhard, K. zu M\"unster, M. Knap, E. Arrigoni, and W. von der Linden,
Phys. Rev. A \textbf{85}, 053644 (2012).

\bibitem{Giamarchi2004}
T.~Giamarchi, \textit{Quantum Physics in One Dimension, International Series of Monographs on Physics}, Vol. 121 (Oxford University Press, Oxford, 2004).

\bibitem{Bissbort:2012}
U. Bissbort, F. Deuretzbacher, and W. Hofstetter, Phys. Rev. A {\bf 86}, 023617 (2012).



\bibitem{Lacki:2013}
 M. Lacki, D. Delande, and J. Zakrzewski, New J. Phys. {\bf 15}, 013062 (2013).


\bibitem{Major:2014}
 J. Major, M. Lacki, and J. Zakrzewski, Phys. Rev. A {\bf 89}, 043626 (2014). 


\bibitem{Chatterjee:2020}
B. Chatterjee, J. Schmiedmayer, C. L\'eveque, and A. U. J. Lode,
arXiv:1904.03966.

\bibitem{Duric:2016}
T.~\DJ uri\'c, K.~Biedro\'n, and J.~Zakrzewski, 
Phys. Rev. B \textbf{95}, 085102 (2017).

\bibitem{Wall2013}
M. L. Wall, and L. D. Carr,
New J. Phys. \ \textbf{15}, 123005 (2013),








\end{thebibliography}
\end{document}